\documentclass[10pt,journal]{IEEEtran} 

\IEEEoverridecommandlockouts
\usepackage{cite}
\usepackage{amsmath,amssymb,amsfonts}
\usepackage{algorithmic}
\usepackage{graphicx}
\usepackage{textcomp}
\usepackage{xcolor}
\usepackage{caption}
\usepackage{subcaption}
\usepackage{array,multirow}
\usepackage[lined,boxed,commentsnumbered,ruled,longend,linesnumbered]{algorithm2e}
\usepackage{enumitem}
\usepackage{multirow} 

\usepackage{tikz}
\usepackage[T1]{fontenc} 
\usepackage{hyperref}
\usepackage{mathtools}

\usepackage{soul}
\usepackage{flushend}

\def\BibTeX{{\rm B\kern-.05em{\sc i\kern-.025em b}\kern-.08em
    T\kern-.1667em\lower.7ex\hbox{E}\kern-.125emX}}

\usepackage{romannum}

\usepackage[acronym]{glossaries}
\makeglossaries

\newacronym{bsc}{BSC}{binary symmetric channel}
\newacronym{bch}{BCH}{Bose, Chaudhuri, and Hocquenghem}
\newacronym{bw}{BW}{bandwidth}
\newacronym{cic}{CIC}{causal influence of communication}
\newacronym{cs}{CS}{central server}
\newacronym{dial}{DIAL}{differentiable inter-agent learning}
\newacronym{fov}{FoV}{field-of-view}
\newacronym{gru}{GRU}{gated recurrent unit}
\newacronym{lc}{LC}{Lipschitz continuous}
\newacronym{iid}{IID}{independent, identically distributed}
\newacronym{LiDAR}{LiDAR}{light detection and ranging}
\newacronym{pd}{PD}{persistence diagram}

\newacronym{nn}{NN}{neural network}
\newacronym{ma}{MA}{multi-agent}
\newacronym{marl}{MARL}{multi-agent reinforcement learning}
\newacronym{mdp}{MDP}{Markov decision processes}
\newacronym{mds}{MDS}{multidimensional scaling}
\newacronym{mmdp}{MMDP}{Multi-agent Markov decision processes}
\newacronym{ml}{ML}{machine learning}
\newacronym{ph}{PH}{persistence homology}
\newacronym{dl}{DL}{deep learning}
\newacronym{jscc}{JSCC}{joint source and channel coding}
\newacronym{mse}{MSE}{mean-squared error}
\newacronym{ntc}{NTC}{nonlinear transform coding}
\newacronym{vr}{VR}{Vietoris-Rips}
\newacronym{sc}{SC}{Semantic Communication}

\newacronym{rial}{RIAL}{reinforced inter-agent learning}
\newacronym{rl}{RL}{reinforcement learning}
\newacronym{sn}{SN}{sensor node}
\newacronym{sp}{SP}{signal processing}
\newacronym{tda}{TDA}{topological data analysis}
\newacronym{tsp}{TSP}{topological signal processing}
\newacronym{vae}{VAE}{variational autoencoder}
\newacronym{i-t}{IT}{information-theoretic}
\newacronym{it}{IT}{information theory}
\newacronym{ib}{IB}{information bottleneck}
\newacronym{nlp}{NLP}{natural language processing}
\newacronym{rd}{RD}{rate-distortion}

\newacronym{tx}{TX}{transmitting node}

\newacronym{rx}{RX}{receiving node}
\newacronym{ae}{AE}{autoencoder}

\newacronym{mqa}{MQA}{average of maximum quantization error}

\newacronym{2d}{2D}{$2$-dimensional}

\usepackage{booktabs} 
\newcommand{\cf}{\emph{cf.}}
\usepackage{hyperref}

\newif\ifshowsps
\showspstrue 

\usepackage{xcolor}

\newtheorem{assump}{Assumption}
\usepackage{todonotes}

\begin{document}

\title{From Raw Data to Structural Semantics: Trade-offs
among Distortion, Rate, and Inference Accuracy\\
}

\author{
\IEEEauthorblockN{Charmin Asirimath,   Chathuranga Weeraddana, Sumudu Samarakoon,~\IEEEmembership{Member,~IEEE}, Jayampathy Ratnayake, and \\ Mehdi Bennis,~\IEEEmembership{Fellow,~IEEE}
}

  \thanks{Charmin Asirimath is with School of Computing, University of Eastern Finland, Finland (e-mail: charmin.pingamage.don@uef.fi).  Chathuranga Weeraddana,  Sumudu Samarakoon, and Mehdi Bennis are with Centre for Wireless Communication, University of Oulu, Finland (e-mail: chathuranga.weeraddana@oulu.fi,  sumudu.samarakoon@oulu.fi, and mehdi.bennis@oulu.fi, respectively). Jayampathy Ratnayake is with the Deptartment of Mathematics, University of Sri Jayewardenepura, Sri Lanka (e-mail: jratnayake@sjp.ac.lk).} 
}

\maketitle

\begin{abstract}

This work explores the advantages of using  \glspl{pd}, topological signatures of raw point cloud data, in a point-to-point communication setting. {\gls{pd} is a structural semantics in the sense that it carries information about the shape and structure of the data}. Instead of transmitting raw data, the transmitter communicates its \gls{pd} semantics, and the receiver carries out inference using the received semantics. We propose novel qualitative definitions for distortion and rate of \gls{pd} semantics while quantitatively characterizing the trade-offs among the distortion, rate, and inference accuracy. Simulations demonstrate that unlike raw data or \gls{ae}-based latent representations, \gls{pd} semantics leads to more effective use of transmission channels, enhanced degrees of freedom for incorporating error detection/correction capabilities, and improved robustness to channel imperfections. For instance, in a binary
symmetric channel with nonzero crossover probability settings, the minimum rate required for \gls{bch}-coded \gls{pd} semantics to achieve an inference accuracy over $80$\% is approximately $15\times$ lower than the rate required for the coded \gls{ae}-latent representations. Moreover, results suggest that the gains of \gls{pd} semantics are even more pronounced when compared with the rate requirements of raw~data. 
\end{abstract}

\glsreset{pd}


\glsresetall

\section{Introduction}~\label{sec:Introduction}

The study of \emph{semantics}, from a communication theory standpoint, dates back to Weaver and Shannon in  1949~\cite{shannon1949mathematical}. Back then, the communication problem was discussed in three levels: Levels A, B, and C, which correspond to the technical problem, semantic problem, and effectiveness problem, respectively \cite[p~4]{shannon1949mathematical}. The technical problem (level A) is based on the mathematical theory of communication introduced by Claude E. Shannon~\cite[p~29]{shannon1949mathematical}, ~\cite{shannon1948mathematical}. 
Subsequently, contemporary related research by Bar-Hillel and Carnap~\cite{carnap1952outline,carnap1954outline,bar1953semantic} argues for the need for new mathematical formulations toward the semantic problem (level B), indicating the limitations of the theories at level A are due to the statistical nature of information. Since then, the literature has not shown much progress for around seven decades in terms of engineering  communication systems until the recent advancements in \gls{ml}. The envisioned futuristic communication system requirements, together with modern \gls{ml} techniques have paved the way for deep insights into computationally extracting semantics from raw data for downstream tasks~\cite{lan2021semantic, strinati20216g,seo2023semantics}. As a result, there has been an unprecedented interest in goal-oriented and semantic communication~\cite{yang2022semantic, qin2021semantic,zhang2022goal, lu2023semantics, gunduz2022beyond,mostaani2022task, getu2024survey}.  

Computational frameworks for extracting semantics from raw data for goal-oriented communications are typically based on \gls{ml} models leveraging information-theoretic frameworks or explicit schemes that yield \emph{structural attributes} of the underlying data. More specifically, \gls{ml}-based semantics are usually extracted based on classic \gls{rd}~\cite{berger1971rate}, and the framework of \gls{ib}~\cite{tishby2000information}. On the other hand, for structure-based techniques, there is a growing interest in integrating topological signatures of data in various downstream tasks~\cite{chazal2021introduction}, which in turn suggests their use in goal-oriented communication. In this respect, \gls{ph} is a powerful tool for capturing topological structures of the data~\cite[\S~5]{chazal2021introduction}. As such, \glspl{pd}~\cite[\S~5.3]{chazal2021introduction} as a summary of \gls{ph}, provide topological signatures of great importance for \gls{ml} applications~\cite{chazal2021introduction,carriere2020perslay, moor2020topological,gabrielsson2020topology,kim2020pllay,reinauer2021persformer,papamarkou2024position,bonis2016persistence,carriere2017sliced,karan2021time,lacombe2018large,skraba2010persistence}.


Unlike \gls{ml}-based semantics, structural semantics such as \glspl{pd} originating from topological primitives have not been explored in terms of their associated trade-offs in a goal-oriented semantic communication system. However, the need for such a study, either theoretical or empirical is of paramount importance and  is the main focus of this paper.

\subsection{Related Work}\label{subsec:Related_Work}

In general, methods for extracting semantic attributes from raw data for goal-oriented communication are divided into two types: those derived from \gls{ml} algorithms~\cite{xie2021deep,zhou2021semantic,wang2022wireless,dai2022nonlinear,torfason2018towards,strouse2017deterministic,liu2022task,shao2021learning,xie2023robust,beck2023semantic,wu2020graph,aguerri2019distributed} and those based on explicit data/information structures~\cite{wang2021performance,wang2017knowledge,ji2021survey,zhou2022cognitive,xiao2022reasoning,zhao2023joint,zhao2022semantic,thomas2022neuro}%
\footnote{There are specific semantic attributes for which the extraction mechanisms may not directly align with the previously mentioned types. For instance, the semantics of `freshness', `relevance', and `value' within the semantic-aware communication architecture proposed for networked control systems~\cite{uysal2022semantic} place a greater emphasis on timeliness and sampling of raw data.}. 
\Gls{ml}-based semantic extraction heavily depends on information theoretic primitives such as classic \gls{rd} and the framework of \gls{ib}, \cf~\cite{xie2021deep, wang2022wireless,zhou2021semantic,torfason2018towards,dai2022nonlinear} and ~\cite{strouse2017deterministic,liu2022task,shao2021learning,xie2023robust,beck2023semantic,wu2020graph,aguerri2019distributed}, respectively. Unlike the \gls{ml}-based semantic extraction methods, techniques for extracting structural semantics of raw data are diverse. However, applications of such techniques in goal-oriented semantic communication are yet to be investigated. 
In the sequel, we discuss in detail the literature related to these two types for extracting semantic
attributes from raw data.

\noindent\textbf{\gls{ml}-based Semantic Extraction:}
%
\\
\indent In formulations based on classic \gls{rd}~\cite{xie2021deep, zhou2021semantic,wang2022wireless,torfason2018towards,dai2022nonlinear}, it is commonplace to study problems using \gls{nlp} (e.g., \cite{xie2021deep, zhou2021semantic}) or \gls{ntc} (e.g., \cite{wang2022wireless,dai2022nonlinear}) that are implemented via \glspl{nn}. 
In certain cases, additional \glspl{nn} are employed to map semantics to channel symbols yielding a \gls{jscc} setup \cite{xie2021deep, wang2022wireless, dai2022nonlinear}. Considered distortions are usually measured in the raw data space by using a cross-entropy model and the rates are measured by using appropriately chosen entropy models.  For instance, maps from images to semantics and semantics to images considered in \cite{torfason2018towards} are based on an \gls{ae}, and the distortion is measured in raw data space. However, unlike the methods in~\cite{xie2021deep, zhou2021semantic,wang2022wireless,dai2022nonlinear}, \cite{torfason2018towards} implements quantization based on dedicated \glspl{nn} that account for the desired rate requirements.


The other class of \gls{ml}-based semantics is based on the \gls{ib} framework originally proposed in~\cite{tishby2000information}. 
In a broader sense, the framework of \gls{ib} enables a minimal representation with a maximal informativeness of some target variable using raw source data. The minimality accounts for complexity, an indication of the rate or the number of bits required to encode the representation. On the other hand, the maximality accounts for the relevance of the representation to the target variable. Consequently, there is a trade-off between informativeness and rate requirements of the resulting representation. Recently, there has been a growing interest in these semantics within the context of goal-oriented communication \cite{strouse2017deterministic,liu2022task,shao2021learning,xie2023robust,beck2023semantic,wu2020graph,aguerri2019distributed}. 

It is worth highlighting that \gls{ml}-based models~\cite{xie2021deep,zhou2021semantic,wang2022wireless,dai2022nonlinear,torfason2018towards,strouse2017deterministic,liu2022task,shao2021learning,xie2023robust,beck2023semantic,wu2020graph,aguerri2019distributed} intrinsically characterize the trade-offs among rate, distortion, and informativeness, since the models are governed by underlying information theoretic frameworks, such as classic \gls{rd} and \gls{ib} frameworks.

\noindent\textbf{Data/Information Structures-based Semantic Extraction:}
\\
\indent Extracting structural semantics from data is a multifaceted process that involves a combination of topology~\cite{chazal2021introduction}, graph theory~\cite{ji2021survey}, clustering~\cite{gan2020data}, dimensionality reduction methods~\cite{bishop2006pattern}, logical formulations~\cite{davis2017logical}, and causal structures~\cite{scholkopf2021toward}. However, the research toward investigating the potential benefits of the choice of structural semantics is still at a very early stage~\cite[\S~II-A]{getu2024survey}.  
%
%
As pointed out in \cite[\S~II-B]{gunduz2022beyond}, knowledge graphs have been used in recent literature highlighting their abilities to extract semantics from raw data, \cf~\cite{wang2021performance,wang2017knowledge,ji2021survey,zhou2022cognitive}. Another graph-based representation of raw data that is closely related to the knowledge graph is proposed in~\cite{xiao2022reasoning}. Authors in~\cite{zhao2023joint} and \cite{zhao2022semantic} have used a mechanism to map raw data into simplicial complexes, an algebraic structure, to be used in inference tasks. The authors of~\cite{thomas2022neuro} present a mechanism for extracting causal structures in the context of goal-oriented semantic communication. 

It is worth pointing out that, in contrast to the \gls{ml}-based semantic extraction, trade-offs among rate, distortion, and informativeness have not been investigated in the case of structural semantics~\cite{wang2021performance,wang2017knowledge,ji2021survey,zhou2022cognitive,xiao2022reasoning,zhao2023joint,thomas2022neuro}, despite their importance in goal-oriented communication system design~standpoint.



\subsection{Motivation and Contribution}

Similar to the structural semantics proposed in \cite{wang2021performance,wang2017knowledge,ji2021survey,zhou2022cognitive,xiao2022reasoning,zhao2023joint,zhao2022semantic,thomas2022neuro}, topological signatures such as \glspl{pd} have shown unprecedented potential in modern \gls{ml} application domains, but have not been considered so far in semantic communication \cite{chazal2021introduction,carriere2020perslay, moor2020topological,gabrielsson2020topology,kim2020pllay,reinauer2021persformer,papamarkou2024position,bonis2016persistence,carriere2017sliced,karan2021time,lacombe2018large,skraba2010persistence}. The proliferation of \glspl{pd} in application domains is emphasized further by the recent research conducted to develop dedicated \gls{ml} architectures for handling \glspl{pd} as inputs \cite{carriere2020perslay, moor2020topological,gabrielsson2020topology,kim2020pllay,reinauer2021persformer}.  

However, the potential of \glspl{pd} as structural semantics in the paradigm of goal-oriented communication is yet to be explored. In this respect, there is a need to investigate trade-offs among \emph{distortion}, \emph{rate requirement}, and \emph{inference accuracy} in a system where \glspl{pd} are used as semantics, which we refer to as \gls{pd} semantics. As such, the key contributions of the paper are summarized~below:
\begin{itemize}
    \item \emph{Benefits of \gls{pd} Semantics}: Conduct a study to explore the advantages of using \glspl{pd} as 
 structural semantics over raw data and \gls{ae}-latent representations in a point-to-point goal-oriented communication setting.
    \item \emph{Communication of Semantics}: Propose strategies for transmitting \gls{pd} semantics from a transmitter (e.g., sensor) to a receiver (e.g., a central node). 
    \item \emph{Qualitative Definitions}: Develop qualitative definitions for semantic distortion and rate within the scope of \glspl{pd}.
    \item \emph{Quantitative Characterization}: Characterize quantitatively the trade-offs among the distortion, the rate, and inference accuracy for a downstream task of interest.
    \item \emph{Empirical Evidence}: Empirically demonstrate that \gls{pd} semantics offer significant advantages over raw data and \gls{ae}-latent representations including:
    \begin{itemize}
     \item Effective use of transmission channels.
     \item Provisioning additional error detection/correction capabilities for enhanced robustness against communication channel impairments.
     \item Robustness of the inference to changes in the distortion and the rate.
    \end{itemize}
\end{itemize}

\subsection{Notation}

Normal font lowercase letters $x$, bold font lowercase letters~$\boldsymbol{x}$, and calligraphic font $\mathcal{X}$ represent scalars, vectors, and sets, respectively. 
The transpose of a vector $\boldsymbol{x}$ is represented by $\boldsymbol{x}^\textrm{T}$.
The sets of real numbers, natural numbers,  and real $n$-vectors are denoted by $\mathbb{R}$, $\mathbb{N}$, and $\mathbb{R}^n$, respectively.
We denote the Euclidean distance and maximum norm by $||\boldsymbol{x}||_2$ and $||\boldsymbol{x}||_\infty$, respectively.  
The nonnegative orthant  ${\mathbb R}^n_{+}$ is the set of points with nonnegative components, i.e., ${\mathbb R}^n_{+} = \{\boldsymbol{x} \in \mathbb{R}^n  \ | \ x_i \geq 0, i=0, \ldots,n \}$. 
The simplex spanned by a list of vectors $\boldsymbol{x}_0,\ldots, \boldsymbol{x}_n$ is denoted by $[\boldsymbol{x}_0,\ldots, \boldsymbol{x}_n]$.

\subsection{Organization of the Paper}
The rest of the paper is organized as follows. The related background is concisely presented in Section~\ref{sec:background}. The system model and problem statement are presented in Section~\ref{sec:System-Model-and-Problem-Statement}. In Section~\ref{sec:Transmission-of-PD-Semantics}, we present a novel approach for transmitting \gls{pd} semantics. Section~\ref{sec:Semantic-Distortion-and-Rate} defines the notion of semantic distortion and  semantic rate, which are used in our subsequent empirical characterizations and associated trade-offs. 
In Section~\ref{sec:Numerical-Results}, empirical characterizations of trade-offs and their implications are presented. Lastly,
Section~\ref{sec:Conclusion} concludes the paper.

\section{Background}\label{sec:background}


In this section, we outline the mathematical concepts and tools used in this paper for generating \gls{pd} semantics of raw point cloud data, \cf~\S~\ref{subsec:Point-Clouds} and \ref{subsec:An-Overview-of-PH}. Moreover, a machinery for adapting \glspl{pd} as inputs for \glspl{nn} is presented in \S~\ref{subsec:Learning-PD-Vectorizations}.

\subsection{Point Clouds}~\label{subsec:Point-Clouds}
A point cloud $\mathcal{G}$ is a set of points that reside in an ambient space, usually the Euclidean space, while each point in the point cloud encodes certain information about the underlying space. From an engineering point of view, sensor technologies including Light Detection and Ranging~(LiDAR), cameras, and structured light cameras are capable of generating such point clouds. In the technical problem (level A) systems, it is commonplace that such sensors routinely transmit their point cloud to a remote server or receiver.
However, transmitting the entire point cloud requires significant resources. In contrast, we propose a novel approach based on communicating a topological signature that encodes topological attributes underlying the point cloud, allowing more efficient communication for solving downstream tasks. 






\subsection{An Overview of \gls{ph}}~\label{subsec:An-Overview-of-PH}
In this subsection, we introduce the necessary mathematical tools for generating a signature that encodes the topological structure of the point cloud. For more details, we refer the reader to~\cite{edelsbrunner2022computational}.


\subsubsection{Simplices and Simplicial Complexes}~\label{subsubsec:Simplices-and-Simplicial-Complexes} 
We start by first introducing the mathematical object called 
 \textit{simplices}, which are generalizations of the notion of a triangle to an arbitrary dimension. Loosely speaking, a $0$-simplex is a point in Euclidean space, a $1$-simplex is a line segment, a $2$-simplex is a filled triangle, a $3$-simplex is a filled tetrahedron, and the concept is generalized similarly for higher order simplices. Formally, 
a $k$-simplex denoted
$\sigma$ is defined as 
\begin{equation}~\label{eq:simplex}
\sigma=\left\{\sum_{i=0}^k \theta_i \boldsymbol{u}_i \ \Bigg | \ \sum_{i=0}^k \theta_i=1, \theta_i \geq 0, i=0,\ldots,k \right\},
\end{equation}
where $\boldsymbol{u}_0,\ldots, \boldsymbol{u}_k$ are affinely independent points in $d$ dimensional Euclidean space $\mathbb{R}^d$, and thus $k \leq d$. The set $\mathcal{U}=\{ \boldsymbol{u}_0, \ldots, \boldsymbol{u}_k \}$ is usually referred to as the vertex set. Moreover, a face of a simplex $\sigma$  whose vertex set is $\mathcal{U}$ are the simplices based on subsets of $\mathcal{U}$. 



Simplicial complexes are based on simplices of the form (\ref{eq:simplex}). Strictly speaking, a simplicial complex $\mathcal{K} \subset \mathbb{R}^n$ is a collection of simplices~[\cf~\S~\ref{eq:simplex}] that satisfy the conditions: $1)$ any face of a simplex of $\mathcal{K}$ is a simplex of $\mathcal{K}$ and $2)$ the intersection of any two simplices of 
$\mathcal{K}$ is either empty or a common face of the two simplices. 


\subsubsection{Filtration}~\label{subsubsec:Filtration} 
A filtration is a sequence of simplicial complexes ($\mathcal{K}_0, \mathcal{K}_1,\dots$) that is necessary for generating the intended topological signature.
More specifically, a filtration of a simplicial complex $\mathcal{K}$ is a nested sequence of simplicial complexes $\emptyset \subset \mathcal{K}_0 \subset \mathcal{K}_1 \subset \cdots \subset \mathcal{K}_m=\mathcal{K}$. 

One can build on top of a given point cloud different filtrations, such as \gls{vr}, Čech, and Witness~\cite{de2004topological} among others, which in turn are used to determine the underlying topological structure. We place a major emphasis on the \gls{vr} filtration since we use it in our subsequent developments. The rule for building the \gls{vr} complex from the point cloud $\mathcal{G}$ relies on a positive scalar~$\gamma$. More specifically, given $\gamma > 0$, the \gls{vr} complex, denoted 
$V(\mathcal{G},\gamma)$ with vertex set $\mathcal{G}$, is defined by the condition 
\begin{multline}~\label{eq:VR}
    [\boldsymbol{g}_0,\boldsymbol{g}_1, \ldots, \boldsymbol{g}_n ] \in V(\mathcal{G},\gamma) \iff || \boldsymbol{g}_i-\boldsymbol{g}_j ||_2 \leq \gamma \\ \forall (i,j)\in \{0,\ldots,n \},
\end{multline}
where $[\boldsymbol{g}_0,\boldsymbol{g}_1, \ldots, \boldsymbol{g}_n ]$ is a face of $ V(\mathcal{G},\gamma)$, whose vertex set is $\{ \boldsymbol{g}_0, \boldsymbol{g}_1, \ldots, \boldsymbol{g}_n\} \subset \mathcal{G}$.  For $\gamma_0 < \gamma_1 <\cdots <\gamma_t$, 
\begin{equation}\label{eq:VR-Filtration}
   V(\mathcal{G},\gamma_0) \subseteq  V(\mathcal{G},\gamma_1) \subseteq \cdots \subseteq V(\mathcal{G},\gamma_t) 
\end{equation}
is held. 

\subsubsection{\gls{ph} and \gls{pd}}~\label{subsubsec:Persistent-Homology-and Persistence-Diagrams}
Based on the filtration, how topological features, such as connected components, loops, and voids, appear and disappear can be computed. These are essentially \emph{structural} properties of the underlying point cloud $\mathcal{G}$.
\begin{figure*}[t!]
\centering
\includegraphics[width=0.8\textwidth]{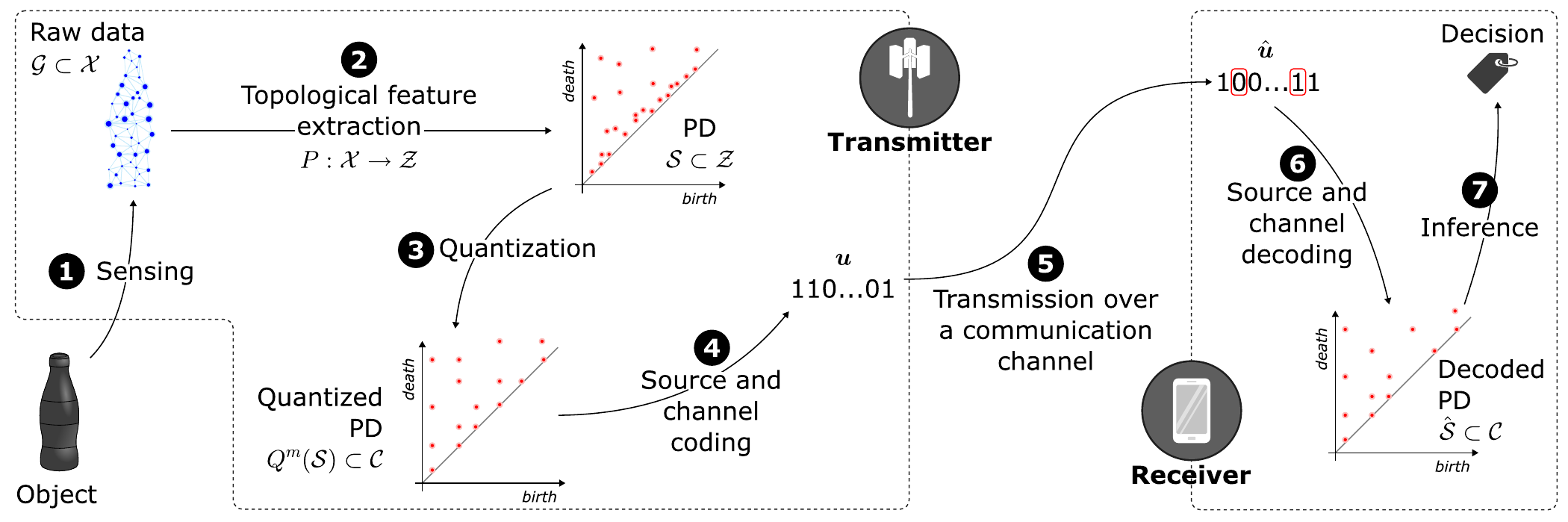}
\caption{System model highlighting different stages.}
\label{fig:Sys-Model}
\end{figure*}
Loosely speaking, the machinery for computing such topological features and their evolution over different spatial resolutions $\gamma$ of the filtration is called \gls{ph}. \gls{ph} gives rise to \gls{ph} groups, one for each non-negative integer $p$, denoted by $\mathcal{H}_p$. For example, $1$-dimensional \gls{ph} group $\mathcal{H}_1$ tracks the evolution of loops that appear in the filtration process. Associated with \gls{ph} of a point cloud is the \gls{pd}, a topological signature that summarizes the evolution of the topological features. The horizontal axis of  \gls{pd} corresponds to the birth time of topological features and the vertical axis represents the death time. 
Hence, \gls{pd} $\mathcal{S}$ is a multi-set of points in the open halfplane that represent the births (in $x$-dimension) and deaths (in $y$-dimension) of all homology classes.

One can define a metric on the space of \glspl{pd} for solving \gls{ml} tasks. In this work, we  measure the distance between two \glspl{pd} $\mathcal{S}_1$ and $\mathcal{S}_2$ by using the  bottleneck distance defined as 
\begin{equation}~\label{eq:Bottleneck}
    W( \mathcal{S}_1,\mathcal{S}_2)_\infty=\inf_{\eta: \mathcal{S}_1 \rightarrow \mathcal{S}_2} \sup_{\boldsymbol{s} \in \mathcal{S}_1} ||\boldsymbol{s}-\eta(\boldsymbol{s})||_{\infty},
\end{equation}
where $\eta$ ranges over all bijections $\mathcal{S}_1$ to $\mathcal{S}_2$.




\subsection{Learning \gls{pd} Vectorizations}~\label{subsec:Learning-PD-Vectorizations}
Despite the potential of using  \glspl{pd} in \gls{ml} applications, the space of \glspl{pd} faces several challenges. For instance, different \glspl{pd} can contain varying numbers of points, and basic operations, such as addition and scalar multiplication, are not
well-defined in the space of \glspl{pd}~\cite{carriere2020perslay}. This makes it challenging to use standard vector space operations in the space of \glspl{pd}, which are integral to many \gls{dl}~algorithms. 

To overcome these challenges, we leverage notions from \gls{tda}, seeking vectorization methods to associate a set of \glspl{pd} with a set of vectors~\cite{carriere2020perslay, kim2020pllay}.  Among these methods, we utilize $\texttt{PersLay}$, a neural network layer for learning  \gls{pd} vectorizations~\cite{carriere2020perslay}. Given a \gls{pd} $\mathcal{S}$, the $\texttt{PersLay}$ is defined through the following equation:
\begin{equation}~\label{eq:PersLay}
    \texttt{PersLay}(\mathcal{S}) \triangleq \texttt{op}( \{w(\boldsymbol{s})\phi(\boldsymbol{s} ) \}_{\boldsymbol{s} \in \mathcal{S} } ),
\end{equation}
where $\texttt{op}$ is a permutation invariant operation, $w:\mathbb{R}^2 \rightarrow \mathbb{R}$ is a weight function for the \gls{pd}, and $\phi:\mathbb{R}^2 \rightarrow \mathbb{R}^n$ is a representation function that maps each point of the \gls{pd} to a~vector.



\section{System Model and Problem Statement}\label{sec:System-Model-and-Problem-Statement}




We consider a point-to-point communication setting consisting of a \gls{tx} acting as a sensor and a \gls{rx} operating as a decision-maker, \cf~Figure~\ref{fig:Sys-Model}. \gls{tx} has point cloud observations over a raw observation space, denoted as $\mathcal{X}$, which will be referred to as raw data hereinafter.
Given any raw data observation $\mathcal{G}\subset\mathcal{X}$ associated with an object (e.g., an image), instead of sharing $\mathcal{G}$ or a compression thereof, \gls{tx} first transforms $\mathcal{G}$ into a semantic space $\mathcal{Z}$ to yield a \gls{pd} $\mathcal{S}$ through a mapping ${P}:\mathcal{X}\rightarrow\mathcal{Z}$ where specific \emph{structural characteristics} of the raw data $\mathcal{G}$ are extracted, \cf~stage $1$ of Figure~\ref{fig:Sys-Model}. 
More specifically, we consider $\mathcal{Z}$ to be the space on which \glspl{pd} or in other words, topological summaries of $\mathcal{G}$  are supported. Thus, $\mathcal{Z}$ is simply the open halfplane given by
$\mathcal{Z}=\{\boldsymbol{s}=(b,d)\ | \  b<d \} \subset {\mathbb R}^2$. In the sequel, we refer to topological summaries of the form ${P}(\mathcal{G})=\mathcal{S}\subset\mathcal{Z}$ as \emph{PD semantics}. Then \gls{tx} transmits respective \gls{pd} semantics to \gls{rx} over a communication channel.

Upon reception of the \gls{pd} semantics, \gls{rx} solely relies on $\mathcal{S}\subset\mathcal{Z}$ for inference. We assume that the downstream task depends upon the output of an inference machinery whose inputs are subsets of $\mathcal{Z}$ carrying structural information of the raw data $\mathcal{G}$, \cf~Figure~\ref{fig:Sys-Model}.
The inference machinery is trained and deployed at \gls{rx}~\footnote{Even though we consider \glspl{pd} as data for training the inference machinery, we must be careful to choose a suitable notion of semantics for our end-task.}. 
More specifically, we consider a \gls{nn} classifier as our inference machinery based on the framework of PersLay~\cf~\S~\ref{subsec:Learning-PD-Vectorizations}. 

It is worth noting that, communicating \gls{pd} semantics $\mathcal{S}$ instead of raw data $\mathcal{G}$ enables an effective representation of the underlying structure of the raw data, yielding an efficient use of the communication channel. As such, we consider a sequence of raw data $\{\mathcal{G}_i\}_{i\in\mathbb{N}}$, that is mapped by \gls{tx} to yield a sequence of \gls{pd} semantics $\{\mathcal{S}_i\}_{i\in\mathbb{N}}$, which in turn are communicated to \gls{rx}, where $\mathcal{S}_i={P}(\mathcal{G}_i)$ for all $i$. 
Given the sequence $\{\mathcal{G}_i\}_{i\in\mathbb{N}}$ of raw data and corresponding sequence $\{\mathcal{S}_i\}_{i\in\mathbb{N}}$ of \gls{pd} semantics, we consider the following problems:
\begin{enumerate}[label={\bfseries P\arabic*}]
\item How to transmit \gls{pd} semantics and quantify their associated semantic distortions and semantic information rate?~\label{P1} 
\item What trade-offs exist between semantic distortion and the semantic rate?~\label{P2} 
\item What trade-offs exist between semantic distortion and inference accuracy?~\label{P3} 
\item What trade-offs exist between inference accuracy and semantic rate?~\label{P4}  
\item What are the implications of such trade-offs from a design point of view?~\label{P5} 
\end{enumerate}
Our approach is to glean insights and understanding into the above problems through the use of simple, instructive arguments, and models. Subsequent formulations in \S~\ref{sec:Transmission-of-PD-Semantics} and \S~\ref{sec:Semantic-Distortion-and-Rate} provide bases for much of our developments for the problem~\ref{P1} above.  These developments, together with the inference machinery, characterize the trade-offs and their implications [\cf~problems~\ref{P2}-\ref{P5}] as discussed in \S~\ref{sec:Numerical-Results}. 



\section{Transmission of \gls{pd} Semantics}\label{sec:Transmission-of-PD-Semantics}

We start by recalling that a \gls{pd} is a multiset supported on the open halfplane 
$\mathcal{Z}$~\cf~\S~\ref{sec:System-Model-and-Problem-Statement}. Given an arbitrary raw data $\mathcal{G}$ from an application domain under consideration, let $\mathcal{S}=\{\boldsymbol{s}_1,\ldots,\boldsymbol{s}_{N_{\mathcal{S}}}\}$ denote the corresponding \gls{pd}, where $N_{\mathcal{S}}$ represents the number of topological features captured during the filtration process, \cf~\S~\ref{subsubsec:Filtration}. Thus, the problem of transmitting \gls{pd} semantics is directly linked with the problem of transmitting each element $\boldsymbol{s}_n$, $n\in\{1,\ldots,N_{\mathcal{S}}\}$, of $\mathcal{S}$, which are simply vectors in ${\mathbb R}^2$. 
To this end, we consider the application of uniform \gls{2d} vector quantization similar to high-rate entropy-coded quantization~\cite[\S~3.5]{gallager2008principles}.
Though adopting more sophisticated schemes such as vector quantization based on the Lloyd–Max algorithm can be easily carried out, it is beyond the scope of this work, and thus, excluded from the analysis.

 \begin{figure*}[!t]
 \centering
 \includegraphics[width=0.8\textwidth]{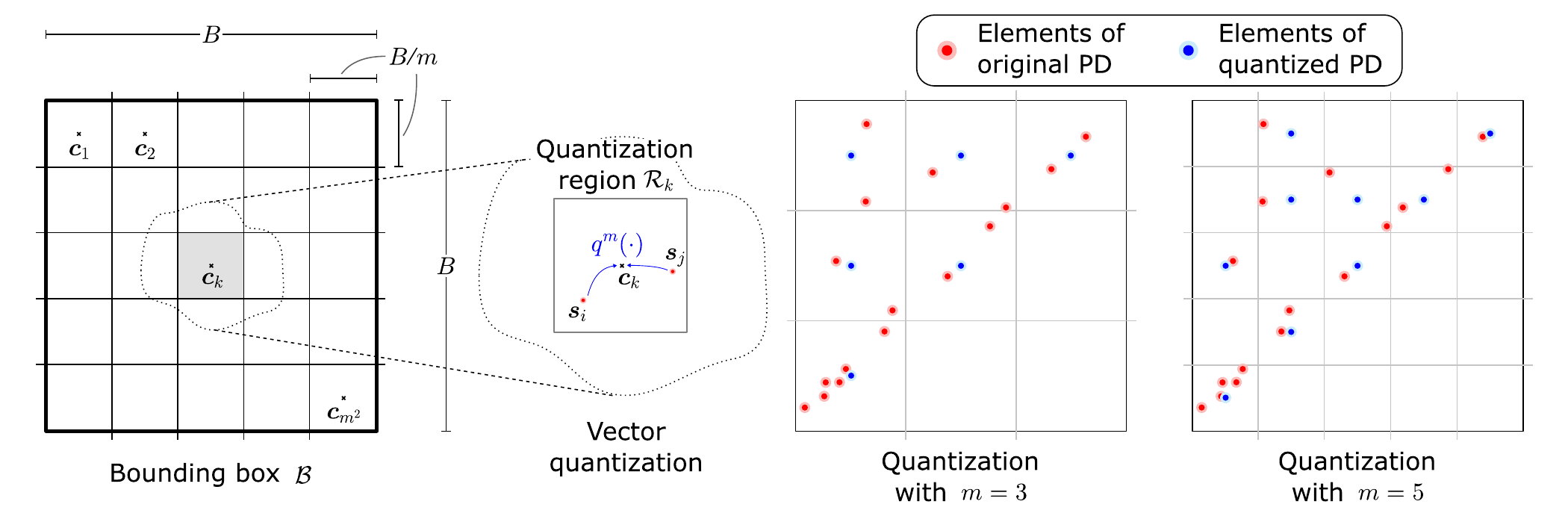}
\caption{Uniform \gls{2d} vector quantization.}
\label{fig:Uniform-2-Dimensional-Vector-Quantization}
  \end{figure*}

\subsection{Quantization of \gls{pd} Semantics}\label{subsec:Quantization-of-PD-Semantics}

Note that the set $\mathcal{S}_i$ itself is the semantics that carries structural properties of a given raw data $\mathcal{G}_i$. Since the elements of $\mathcal{S}_i$ are vectors in the nonnegative orthant ${\mathbb R}^2_{+}$, we model the semantic input to the quantizer as a sequence of \gls{2d} analog random variables. In particular, the input to the quantizer will be $\{\mathcal{S}_i\}_{i\in\mathbb N}$, where $\mathcal{S}_i=\{\boldsymbol{s}_{i,1},\ldots,\boldsymbol{s}_{i,N_{\mathcal{S}_i}}\}$ is the \gls{pd} semantics of raw data $\mathcal{G}_i$, $\boldsymbol{s}_{i,j}\in\mathbb{R}^2$, and $N_{\mathcal{S}_i}$ is a finite integer specific to $\mathcal{S}_i$. Moreover, the elements of $\mathcal{S}_i$ for all $i$ are assumed to be
\gls{iid} according to a compactly supported probability density function $f_{\boldsymbol{s}}:\mathbb{R}^2\rightarrow \mathbb R$. Thus, we have the following assumption about the elements of $\mathcal{S}_i$.
\begin{assump}\label{Assumption:Bounding-Box-of-Semantics}
Elements of $\mathcal{S}_i$ for all $i$ are uniformly bounded. That is, there exists a finite $B$ such that for all $i$,
\begin{equation*}
    \max_{n\in\{1,\ldots, N_{{\mathcal{S}}_i}\}}||\boldsymbol{s}_{i,n}||_{\infty} \leq B.
\end{equation*}
\end{assump}
Note that from a practical point of view, the above assumption is not a restriction. For example, it is common in almost all applications that the raw data $\mathcal{G}_i$ is compact. This in turn enables one to come up with a positive scalar $\gamma_t\geq B$ [\cf~\eqref{eq:VR-Filtration}] beyond which no further topological features will appear, and thus $\mathcal{Z}$ becomes compact.

For all $i$ and $j\in\{1,\ldots,N_{\mathcal{S}_i}\}$, the quantizer maps the elements $\boldsymbol{s}_{i,j}$ of $\mathcal{S}_i$, which are random vectors in ${\mathbb R}^2$ into \emph{discrete} \gls{2d} random vectors $\boldsymbol{q}_{i,j}$, so that the set $\mathcal{C}$ of representation points for each $\boldsymbol{q}_{i,j}$ is a \emph{finite} set with cardinality~$m^2$, i.e., $\boldsymbol{q}_{i,j}\in\mathcal{C}=\{\boldsymbol{c}_1,\ldots,\boldsymbol{c}_{m^2}\}$, \cf~Figure~\ref{fig:Uniform-2-Dimensional-Vector-Quantization}. In particular, the semantic space $\mathcal{Z}$ is enclosed inside the box $\mathcal{B} = \{\boldsymbol{b}\in\mathbb{R}^2 \ | \ [0 \ 0]^{\textrm{T}}\leq \boldsymbol{b}\leq [B \ B]^{\textrm{T}}\}$, and each side of the box is divided into $m$ equal intervals. We refer to $m$ as the \emph{bins per dimension} considered for quantization. Consequently, $\mathcal{B}$ is split into $m^2$ identical and mutually exclusive square-shaped regions. The square-shaped regions are simply the \emph{quantization regions}, which we denote by $\mathcal{R}_k$, $k\in\{1,\ldots,m^2\}$. Let $\mathcal{R}$ be the set of quantization regions, i.e., $\mathcal{R}=\{\mathcal{R}_1,\ldots,\mathcal{R}_{m^2}\}$. The geometrical centers of the quantization regions in $\mathcal{R}$ are considered to be the \emph{representation points} $\boldsymbol{c}_k$, $k\in\{1,\ldots,m^2\}$. 
Formally, given \gls{pd} semantics $\mathcal{S}=\{\boldsymbol{s}_1,\ldots,\boldsymbol{s}_{N_\mathcal{S}}\}$, a finite sequence of analog \gls{2d} vectors in $\mathcal{Z}$, the quantizer is viewed as a map $Q^m:\otimes_{n=1}^{N_{\mathcal{S}}}{\mathbb R}^2\rightarrow \otimes_{n=1}^{N_{\mathcal{S}}}{\mathbb R}^2$ that takes as input the set $\mathcal{S}$ and outputs a set of discrete \gls{2d} vectors in $\mathcal{C}$ given by
\begin{equation}\label{eq:Quantization-Map}
 Q^m(\mathcal{S}) = \{q^m(\boldsymbol{s}_1),\ldots,q^m(\boldsymbol{s}_{N_\mathcal{S}})\},
\end{equation}
where $q^m:{\mathbb R}^2\rightarrow{\mathbb{R}}^2$ is a \gls{2d} vector quantizer defined as
\begin{equation}\label{eq:Quantization-Map-Sub}
 \boldsymbol{s}\in\mathcal{R}_k \implies   q^m(\boldsymbol{s}) = \boldsymbol{c}_k.
\end{equation}
Here, the superscript $m$ is used to indicate the number of bins per dimension considered for the quantization.

\subsection{Transmission of \gls{pd} Semantics and Reconstruction at the~Receiver}\label{subsec:Transmission-of-PD-Semantics}

The representation points $\boldsymbol{c}_k$, $k\in\{1,\ldots,m^2\}$ are mapped into an $m^2$-symbol alphabet $\mathcal{A}=\{1,\ldots, m^2\}$, which in turn enables the transmitter to communicate \gls{pd} semantics through a classic digital communication system \cf~Figure~\ref{fig:Sys-Model}~\cite[\S~3]{gallager2008principles}. In particular, given an observation $\mathcal{S}$, symbols from the alphabet $\mathcal{A}$ that correspond to the quantized \gls{pd} semantics $Q^m(\mathcal{S})$ are encoded into binary digits yielding the related sequence ${\boldsymbol{u}}$ of binary symbols, \cf~stage $4$ of Figure~\ref{fig:Sys-Model}. The sequence ${\boldsymbol{u}}$ is transmitted by \gls{tx} to \gls{rx} through the available communication channel, \cf~stage $5$ of Figure~\ref{fig:Sys-Model}.

%
%
At the receiver, the received binary sequence $\hat{\boldsymbol{u}}$ symbols are converted back to the corresponding \gls{2d} representation points in $\mathcal{C}$ to yield the corresponding received \gls{pd} semantics $\hat{\mathcal{S}}$ of the originally observed \gls{pd} ${\mathcal{S}}$ for all $i\in\mathbb{N}$, \cf~stage $6$ of Figure~\ref{fig:Sys-Model}. Note that ${\boldsymbol{u}}$ and $\hat{\boldsymbol{u}}$ are not necessarily identical unless the communication between \gls{tx} and \gls{rx} is~perfect.


\section{\gls{pd}-Semantic Distortion and Rate}\label{sec:Semantic-Distortion-and-Rate}
In this section, we quantify the distortion and rate of \gls{pd} semantics. The purpose is to use them as a basis for subsequent empirical evaluations of the trade-offs, \cf~\S~\ref{sec:System-Model-and-Problem-Statement}. Finally, analogous definitions for \gls{ae} and raw data are also used in the empirical evaluations as a benchmark.

\subsection{\gls{pd}-Semantic Distortion}\label{subsec:Semantic-Distortion}

Given a quantization map $Q^m$ over the semantic space $\mathcal{Z}$ [\cf~\S\ref{subsec:Quantization-of-PD-Semantics}], let us next give a simple definition for the resulting semantic distortion due to quantization.
The definition for distortion denoted $D_{\textrm{MSE}}^{\textrm{PD}}$ is inspired by the notion of classic \gls{mse} and is given~by
\begingroup
\allowdisplaybreaks
\begin{align}\label{eq:Semantic-Distortion-MSE}
    D_{\textrm{MSE}}^{\textrm{PD}} & = \mathbb{E}_{f_{\boldsymbol{s}}} \ ||\boldsymbol{s}-q^m(\boldsymbol{s})||^2\\ \allowdisplaybreaks 
     & = \int_{\mathcal{B}} f_{\boldsymbol{s}}(\boldsymbol{s}) ||\boldsymbol{s}-q^m(\boldsymbol{s})||_2^2 \ d\boldsymbol{s}\\ \allowdisplaybreaks
     & = \sum_{k=1}^{m^2}\int_{\mathcal{R}_k} f_{\boldsymbol{s}}(\boldsymbol{s}) ||\boldsymbol{s}-\boldsymbol{c}_k||_2^2 \ d\boldsymbol{s}, \allowdisplaybreaks
\end{align}
\endgroup
where the expectation is taken with respect to the distribution~$f_{\boldsymbol{s}}$. The superscript indicates that \gls{pd} semantics are used in the computation. 

It is worth pointing out that distortions can also be associated with the \gls{pd} semantics $\mathcal{S}= \{\boldsymbol{s}_1,\ldots,\boldsymbol{s}_{N_{\mathcal{S}}}\}$. In such a setting, not only  $\boldsymbol{s}_i$s, but also $N_\mathcal{S}$ is modeled as a random variable, \cf~Appendix~\ref{App:Dist-over-S}.

\subsection{\gls{pd}-Semantic Rate}\label{subsec:Semantic-Information}

We start by defining a notion of semantic information within the quantization setting given in \S~\ref{sec:Transmission-of-PD-Semantics}. More specifically, semantic information is quantified by considering the entropy of the output of the quantizer $q^m$ [\cf~\eqref{eq:Quantization-Map-Sub}]. In particular, we define by $p_k$ the probability of $\boldsymbol{s}$ residing in region $\mathcal{R}_k$, i.e.,
\begin{equation}\label{eq:Probability-s-in-Rk}
    p_k = \int_{\mathcal{R}_k} f_{\boldsymbol{s}}(\boldsymbol{s}) \ d\boldsymbol{s}, \quad k=1,\ldots,m^2.
\end{equation}
Consequently, the entropy $H$ of the quantizer $q^m$ is given by
\begin{equation}\label{eq:Entropy-of-quantizer-q}
    H(q^m) = - \sum_{k=1}^{m^2} p_k\log_2 \ p_k,
\end{equation}
where the unit is in bits per symbol and the symbols are chosen from the alphabet $\mathcal{A}$, \cf~\S~\ref{subsec:Transmission-of-PD-Semantics}. 
Finally, we define the rate of \gls{pd} semantics as $R^{\textrm{PD}}$ [bits/object], where
\begin{equation}\label{eq:Rate-of-PD-Semantics}
    R^{\textrm{PD}} = M_{\mathcal{S}} H(q^m).
\end{equation}
Note that $M_{\mathcal{S}}={m(m+1)}/{2}$ and is introduced since $p_k$ is nonzero only for $M_{\mathcal{S}}$ quantization regions, \cf~Figure~\ref{fig:Uniform-2-Dimensional-Vector-Quantization}. This is a direct consequence of $f_{\boldsymbol{s}}$ being \emph{zero} over quantization regions lying in the right-bottom triangular space of $\mathcal{B}$, which follows from the definition of \glspl{pd}~\cf~\S~\ref{subsubsec:Persistent-Homology-and Persistence-Diagrams}.





\subsection{Rate and Distortion Definitions for \gls{ae} based latent representations}\label{subsec:Benchmark-ae} 

For comparison, we have also considered an \gls{ae}, whose latent dimension is $2d$, where $d\in\mathbb{N}$. As a result, for a given raw data $\mathcal{G}_i$, the corresponding \gls{ae}-latent representation denoted $\mathcal{A}_i^d$, is given by $\mathcal{A}_i^d=\{\boldsymbol{a}_{i,1},\ldots,\boldsymbol{a}_{i,d}\}$, where $\boldsymbol{a}_{i,j}\in\mathbb{R}^2$. In other words, the raw data $\mathcal{G}_i$ is mapped into a set $\mathcal{A}_i^d$ of \gls{2d} points whose cardinality is $d$.

The developments and definitions considered in \S~\ref{sec:Transmission-of-PD-Semantics}, \S~\ref{subsec:Semantic-Distortion}, and \S~\ref{subsec:Semantic-Information}, are applied in the same manner, except that 
$\mathcal{S}_i$ is replaced by $\mathcal{A}_i^d$, $N_{\mathcal{S}_i}$ is replaced by $N_{\mathcal{A}_i^d}$, $f_{\boldsymbol{s}}$ is replaced by $f^d_{\boldsymbol{a}}$, and ${M}_{\mathcal{S}}$ is replaced by ${M}_{\mathcal{A}}$. Note that we have $N_{\mathcal{A}_i^d}=d$ for all $i$. Moreover, the \gls{2d} points in the latent representations $\mathcal{A}_i^d$ for all $i$ are assumed to be distributed according to the density function $f^d_{\boldsymbol{a}}:\mathbb{R}^2\rightarrow \mathbb R$. Note that ${M}_{\mathcal{A}}$ is the number of quantization regions in which the respective probabilities are nonzero, \cf~\eqref{eq:Probability-s-in-Rk}. It is worth pointing out that, unlike the case with \gls{pd} semantics where $f_{\boldsymbol{s}}$ is guaranteed to be zero over explicit regions in $\mathcal{B}$, $f^d_{\boldsymbol{a}}$ is not necessarily zero in any explicit region. 
Therefore, we have ${M}_{\mathcal{A}}= m^2$. 
Furthermore, without loss of generality, we can choose $\mathcal{B}$ in Assumption~\ref{Assumption:Bounding-Box-of-Semantics} to be the same even with the latent representations $\{\mathcal{A}_i^d\}_{i\in\mathbb{N}}$, since the latent space can be uniformly scaled and~translated.

Finally, the resulting distortion definition for the \gls{ae}-latent representations is denoted by $D^{\textrm{AE},d}_{\textrm{MSE}}$. Moreover, the corresponding data rate is denoted by $R^{\textrm{AE},d}$.

\subsection{Rate and Distortion Definitions for Raw Data}\label{subsec:Benchmark} 
%
%

Analogously, developments and definitions related to the raw data are derived and outlined below without further details since they are clear from the context. The sequence of raw data $\{\mathcal{G}_i\}_{i\in\mathbb{N}}$ is considered with the cardinality of $\mathcal{G}_i$ being $N_{\mathcal{G}_i}$. The density function is denoted by $f_{\boldsymbol{g}}$. The number of quantization regions is denoted by ${M}_{\mathcal{G}}$ which equals $m^2$ similar to the case of ${M}_{\mathcal{A}}$.

As such, the resulting distortion definition for raw data is denoted by $D^{\textrm{raw}}_{\textrm{MSE}}$. Moreover, the corresponding data rate for raw data is denoted by $R^{\textrm{raw}}$.



\section{Empirical Characterization of Trade-Offs and Their Implications}\label{sec:Numerical-Results}


We are now ready to characterize the underlying trade-offs in terms of \gls{pd}-semantic distortion $D_{\textrm{MSE}}^{\textrm{PD}}$, \gls{pd}-semantic rate $R^{\textrm{PD}}$, and inference accuracy which we denote by $A^{\textrm{PD}}$, \cf~problems~\ref{P2}-\ref{P5} in \S~\ref{sec:System-Model-and-Problem-Statement}. Recall that the distortions of \gls{ae}-latent representations and raw data are denoted by $D_{\textrm{MSE}}^{\textrm{AE},d}$ and $D_{\textrm{MSE}}^{\textrm{raw}}$, and the respective rates are denoted by $R^{\textrm{AE},d}$ and $R^{\textrm{raw}}$, respectively, \cf~\S~\ref{subsec:Benchmark-ae}, \S~\ref{subsec:Benchmark}. We further denote by $A^{\textrm{AE},d}$ and $A^{\textrm{raw}}$ the inference accuracy of \gls{ae}-latent representations and raw data based systems, respectively. The empirical setup is based on a point cloud dataset based on the MNIST~\cite{deng2012mnist}, \cf~\S\ref{subsec:Dataset}. Finally, we outline the training and testing environments [\S~\ref{subsec:Training-and-Testing}] followed by the empirical results [\S~\ref{subsec:Experiments}].

\subsection{Dataset and Preprocessing}\label{subsec:Dataset}

Note that the proposed developments are generally applied to point cloud data. Therefore, we first create our point cloud dataset based on MNIST~\cite{deng2012mnist} which compromises images (objects) of handwritten digits with a size of $28 \times 28$ pixels. Given an image from the dataset, each pixel coordinate of the image is modeled by an integer pair $(x,y)$, where $(x,y)$ is an element of the grid $\{(x,y)\in{\mathbb{N}}^2 \ | \ 1 \leq x,y \leq 28 \} \subset \mathbb{R}^2$. Using the dataset, we chose a set $\mathcal{N}$ of $600$ images and used them throughout our experiments. As a preprocessing step for obtaining realizations of raw random elements $\mathcal{G}_i$s, we use images in $\mathcal{N}$. In particular, for each image $i$ in $\mathcal{N}$, we associate the observation $\mathcal{G}_i=\{ (x,y) \ | \ g^i(x,y) \geq 0.70 \}$~\footnote{\label{note1}For notational simplicity, we use the same notation $\mathcal{G}_i$ and $\mathcal{S}_i$ even when denoting an observation of random $\mathcal{G}_i$ and $\mathcal{S}_i$, respectively.}. Here $g^i:(x,y)\mapsto [-1,1]$ is the grayscale of the pixel coordinate $(x,y)$ of image~$i$. Thus, we have the finite sequence $\{\mathcal{G}_i\}_{i\in\mathcal{N}}$ of raw data observations. Moreover, the number of classes~$c$ for classification is considered to be $3$ [i.e., $c=3$] and is determined by the number of loops of the digits. 
More specifically, class label $j$, $j\in\{1,2,3\}$ is denoted by $\mathcal{C}_j$, where $\mathcal{C}_1=\{ 0,6,9 \}$ with each element having one loop,  $\mathcal{C}_2=\{ 8 \}$ with each element having two loops, and $\mathcal{C}_3=\{ 1,2,3,4,5,7 \}$ with each element having no loops~\footnote{We did not consider $4$ as a digit with a loop since the dataset consists of \emph{handwritten} digits. Therefore usually $4$ doesn't contain a loop.}. The number of examples is set to be identical, i.e., $200$ per~class. The raw dataset $\{\mathcal{G}_i\}_{i\in\mathcal{N}}$ and its corresponding labels are available at \cite{ICON_Repo_Data}. Based on the raw data, together with class labels, we train \gls{nn}-based classifiers as we discuss next.

\subsection{Training/Testing, and Empirical Computations}\label{subsec:Training-and-Testing}
Recall that we have the finite sequence $\{\mathcal{G}_i\}_{i\in\mathcal{N}}$ of observations as our raw dataset and corresponding class labels. Associated with each $\mathcal{G}_i$, we have the corresponding observation $\mathcal{S}_i$ of \gls{pd} semantics~\textsuperscript{\ref{note1}}. For generating \gls{ae}-latent representations of dimension $2d$, we train a \gls{nn} model denoted by $\texttt{Enc}^d$  using a commonly used unsupervised training procedure applied to the raw data $\{\mathcal{G}_i\}_{i\in\mathcal{N}}$. Specifically, we have $\texttt{Enc}^d(\mathcal{G}_i)=\mathcal{A}^d_i$ for all $i\in\mathcal{N}$, see Appendix~\ref{App:Para-NN}. Moreover, we have computed realizations of quantized \gls{pd} semantics, quantized \gls{ae}-latent representations, and quantized raw datasets. In particular, for all $i\in\mathcal{N}$, we compute the realizations $\{Q^m(\mathcal{S}_i)\}_{m\in\{10,\ldots,27\}}$ of quantized \gls{pd} semantics, the realizations $\{Q^m(\mathcal{\mathcal{A}}^d_i)\}_{m\in\{10,\ldots,27\}}$ of quantized \gls{ae}-latent representations, and the realizations $\{Q^m(\mathcal{G}_i)\}_{m\in\{10,\ldots,27\}}$ of quantized raw data~\footnote{The $(x,y)$-grid of $\mathcal{G}_i$ $\forall$ $i$ is $28\times 28$. Thus, we let $m\in\{10,\ldots,27\}$.} for testing purposes. When training our \gls{nn} models for classification, we use the sequence of observed \gls{pd} semantics $\{\mathcal{S}_i\}_{i\in\mathcal{N}}$, corresponding \gls{ae}-latent representations $\{\mathcal{A}^d_i\}_{i\in\mathcal{N}}$, and the corresponding observations of raw data~$\{\mathcal{G}_i\}_{i\in\mathcal{N}}$, together with the class labels.


Our training and testing are based on repeating $2$-fold cross-validation \cite[\S~7.10]{hastie2009elements} over the chosen examples for $T=25$ times. First, the set of indices $\mathcal{N}=\{1,\ldots,600\}$ is partitioned randomly into a training index set $\mathcal{N}^{\textrm{train}}_t$ and a testing index set $\mathcal{N}^{\textrm{test}}_t$ with each having a cardinality of $300$. Then, for every $t\in\{1,\ldots,T\}$, \gls{nn} models are trained by considering the sequence of \gls{pd} semantics $\{\mathcal{S}_i\}_{i\in\mathcal{N}^{\textrm{train}}_t}$, \gls{ae}-latent representations $\{\mathcal{A}^d_i\}_{i\in\mathcal{N}^{\textrm{train}}_t}$, and the sequence of raw data $\{\mathcal{G}_i\}_{i\in\mathcal{N}^{\textrm{train}}_t}$. 
The corresponding \gls{nn} models for classification are denoted by, $\texttt{PD}_t$, $\texttt{AE}^d_t$, and $\texttt{Raw}_t$, where the superscript $d$ signifies that the latent dimension of \gls{ae} is $2d$. Details of the models are given in Appendix~\ref{App:Para-NN}.

Testing is conducted by using quantized \gls{pd} semantics, quantized \gls{ae}-latent representations, and quantized raw data. More specifically, for fixed bins per dimension $m$, test examples for corresponding quantized \gls{pd} semantics $\{Q^m(\mathcal{S}_i)\}_{i\in\mathcal{N}^{\textrm{test}}_t}$ are used with the trained model $\texttt{PD}_t$ for all $t\in\{1,\ldots,T\}$. 
As such, an average inference accuracy $\hat A^{\textrm{PD}}$ is computed, which is given by 
\begin{equation}\label{eq:Av-Inf-Acuracy}
    \hat A^{\textrm{PD}} = \textstyle(1/T)\sum_{t=1}^T A^{\textrm{PD}}(m,t),
\end{equation}
where $A^{\textrm{PD}}(m,t)$ is the average error that corresponds to testing $\{Q^m(\mathcal{S}_i)\}_{i\in\mathcal{N}^{\textrm{test}}_t}$ with the model $\texttt{PD}_t$. Similarly, we denote by $\hat A^{\textrm{AE},d}$ and $\hat A^{\textrm{raw}}$ the average errors that correspond to testing $\{Q^m(\mathcal{A}^d_i)\}_{i\in\mathcal{N}^{\textrm{test}}_t}$ with $\texttt{AE}_t^d$ models and $\{Q^m(\mathcal{G}_i)\}_{i\in\mathcal{N}^{\textrm{test}}_t}$ with $\texttt{Raw}_t$ models, respectively.

Note that the probability distribution $f_{\boldsymbol{s}}$ is not readily available in closed form. Therefore, when computing distortions $D_{\textrm{MSE}}^{\textrm{PD}}$ [\cf~\eqref{eq:Semantic-Distortion-MSE}] and $ R^{\textrm{PD}}$ [\cf~\eqref{eq:Probability-s-in-Rk}, \eqref{eq:Entropy-of-quantizer-q}, and \eqref{eq:Rate-of-PD-Semantics}], we use an estimate $\hat{f}_{\boldsymbol{s}}$ of $f_{\boldsymbol{s}}$ based on the re-scaled histogram of \gls{2d} analog vectors $\boldsymbol{s}_{i,1},\ldots,\boldsymbol{s}_{i,N_{\mathcal{S}_i}}$ that constitute $\mathcal{S}_i$ for all $i\in\mathcal{N}_t^\textrm{test}$. More specifically, $\mathcal{B}$ is partitioned into $28\times 28$ mutually exclusive boxes with $\mathcal{R}_{ab}$ denoting the $(a,b)$-th box, and $\hat{f}_{\boldsymbol{s}}$ supported over $\mathcal{B}$ is given by
\begin{equation}\label{eq:empirical-probability-distribition}
  \hat{f}_{\boldsymbol{s}}(\boldsymbol{s}) = \beta \textstyle\sum_{t=1}^T I_{\mathcal{N}_t^{\textrm{test}}}(a,b), \quad \boldsymbol{s}\in\mathcal{R}_{ab},
\end{equation}
where $I_{\mathcal{N}_t^{\textrm{test}}}(i,j)$ is the number of occurrences that $\boldsymbol{s}_{i,j}$ falls within the region $\mathcal{R}_{ab}$ for some $i\in\mathcal{N}_t^{\textrm{test}}$ and $j\in\{1,\ldots,N_{\mathcal{S}_i}\}$. The scalar $\beta$ is simply a scaling factor appropriately chosen so that $\int_{\mathcal{B}}\hat{f}_{\boldsymbol{s}}(\boldsymbol{s}) \ d\boldsymbol{s}=1$. The resulting numerical values of $D_{\textrm{MSE}}^{\textrm{PD}}$ and $ R^{\textrm{PD}}$ are denoted by $\hat D^{\textrm{PD}}_{\textrm{MSE}}$ and $ \hat R^{\textrm{PD}}$, respectively.  
The numerical values $\hat D_{\textrm{MSE}}^{\textrm{AE},d}$ and $ \hat R^{\textrm{AE},d}$ of $D_{\textrm{MSE}}^{\textrm{AE},d}$ and $ R^{\textrm{AE},d}$, respectively, are also computed, by using an estimate $\hat{f}^d_{\boldsymbol{a}}$ of $f^d_{\boldsymbol{a}}$ based on the re-scaled histogram of \gls{2d} analog vectors $\boldsymbol{a}_{i,1},\ldots,\boldsymbol{a}_{i,N_{\mathcal{A}^d_i}}$ that constitute $\mathcal{A}^d_i$ for all $i\in\mathcal{N}$. Empirical evaluations of the distortion $\hat D_{\textrm{MSE}}^{\textrm{raw}}$ and the rate $ \hat R^{\textrm{raw}}$  corresponding to raw data are analogously computed.




\begin{figure}[!t] 
 \centering
 \includegraphics[width=\linewidth]{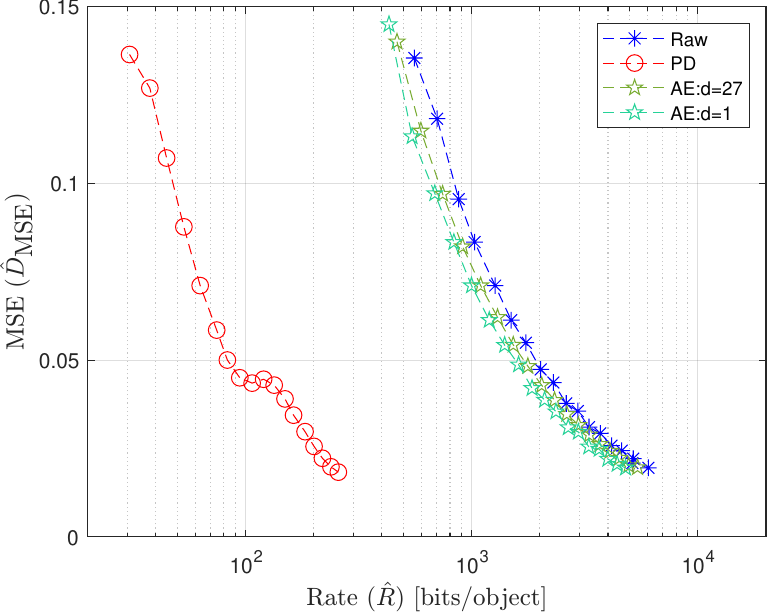}
   \caption{Empirical trade-off between $ \hat D_{\textrm{MSE}}$ and $\hat R$.}
   \label{fig:Trade-offs-between-Semantic-Distortion-and-Rate}
\end{figure}

\subsection{Experiments}\label{subsec:Experiments}

\begin{figure*}[!t] 
 \centering
   \begin{subfigure}{0.326\textwidth}
     \includegraphics[width=\linewidth,height=5cm]{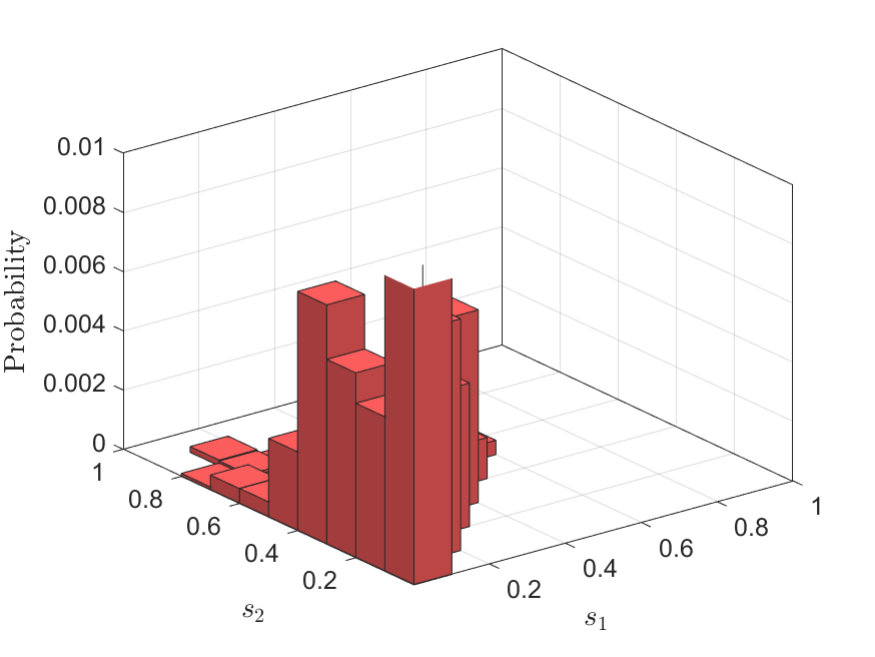}
       \caption{$\hat{f}_{\boldsymbol{s}}$ for $m=10$}
       \label{fig:Dist-PD1}
   \end{subfigure}
\hfill 
   \begin{subfigure}{0.326\textwidth}
       \includegraphics[width=\linewidth,height=5cm]{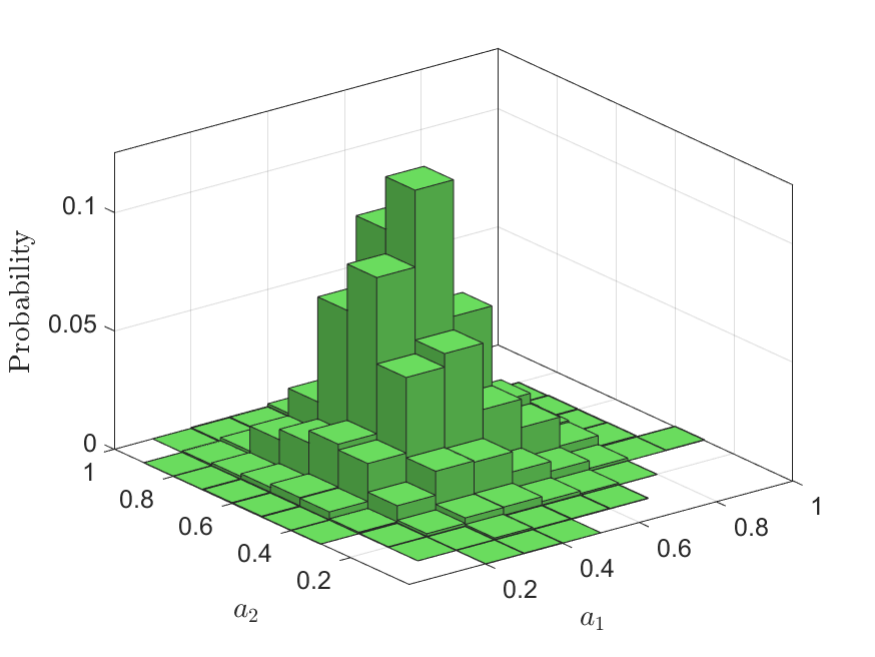}
       \caption{$\hat{f}^{27}_{\boldsymbol{a}}$ for $m=10$}
       \label{fig:Dist-PD2}
   \end{subfigure}
\hfill 
   \begin{subfigure}{0.326\textwidth}
       \includegraphics[width=\linewidth,height=5cm]{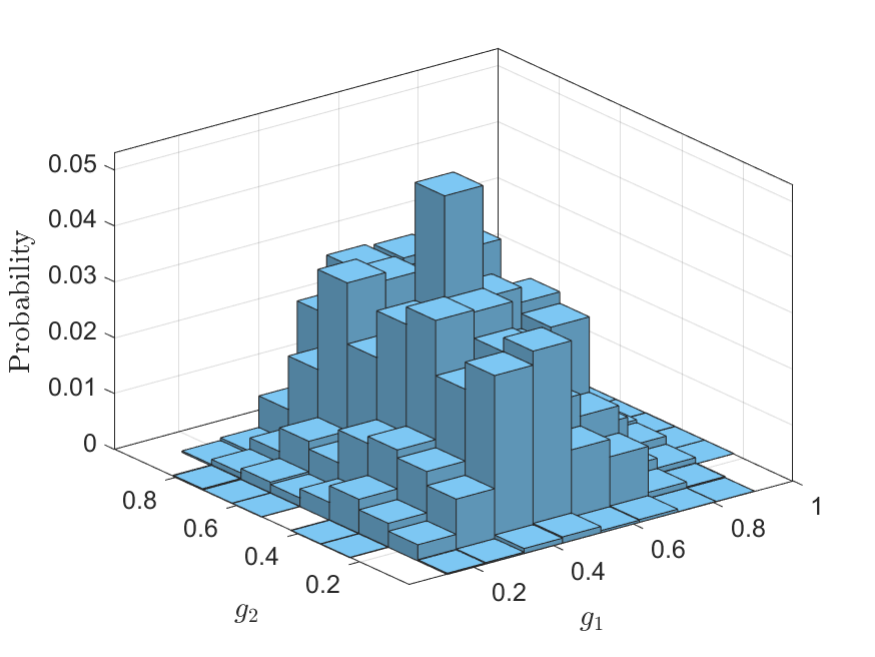}
       \caption{$\hat{f}_{\boldsymbol{g}}$ for $m=10$}
       \label{fig:Dist-PD3}
   \end{subfigure}

\begin{subfigure}{0.326\textwidth}
     \includegraphics[width=\linewidth,height=5cm]{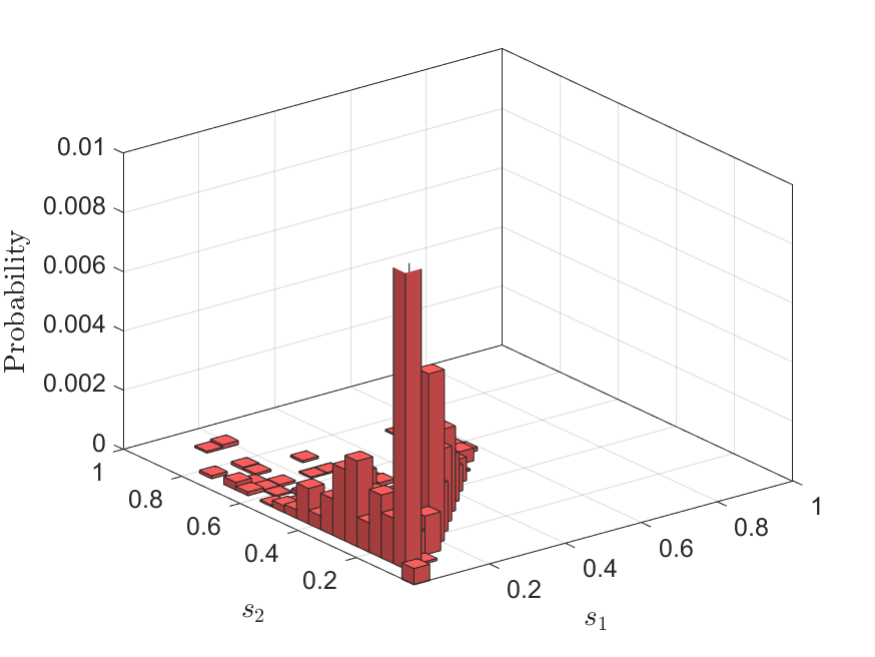}
       \caption{$\hat{f}_{\boldsymbol{s}}$ for $m=24$}
       \label{fig:Dist-PC1}
   \end{subfigure}
\hfill 
   \begin{subfigure}{0.326\textwidth}
       \includegraphics[width=\linewidth,height=5cm]{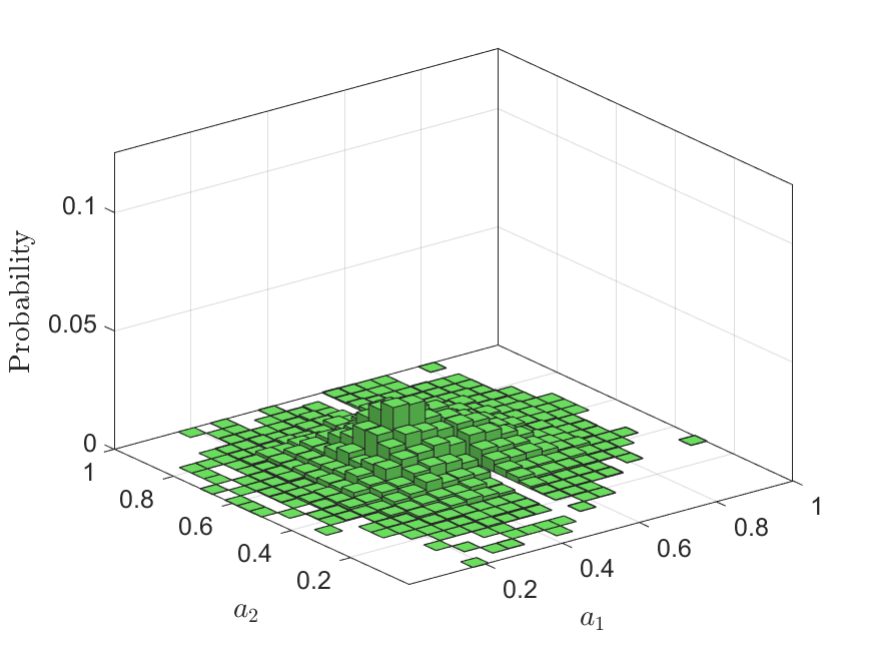}
       \caption{$\hat{f}^{27}_{\boldsymbol{a}}$ for $m=24$}
       \label{fig:Dist-PC2}
   \end{subfigure}
\hfill 
   \begin{subfigure}{0.326\textwidth}
       \includegraphics[width=\linewidth,height=5cm]{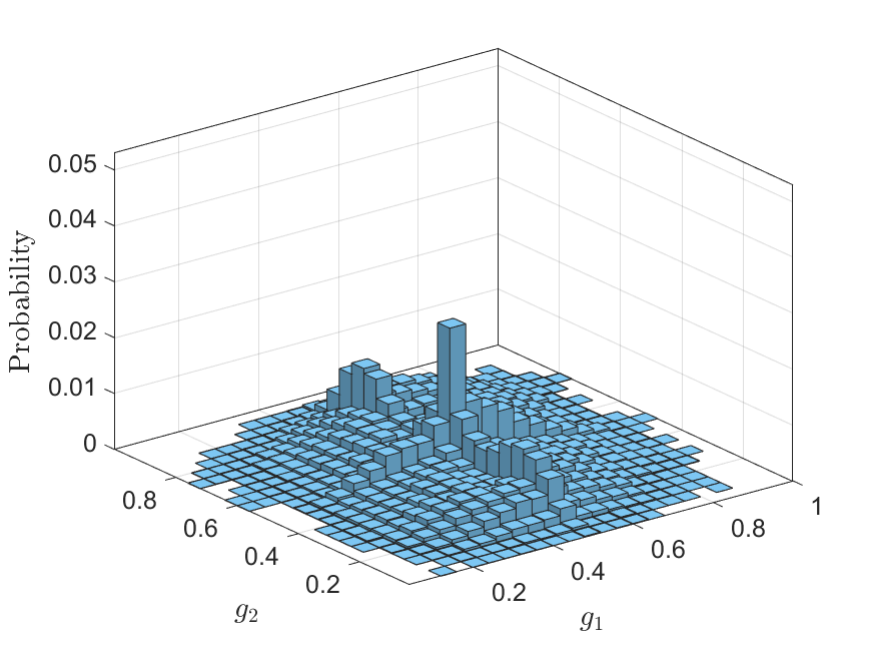}
       \caption{$\hat{f}_{\boldsymbol{g}}$ for $m=24$}
       \label{fig:Dist-PC3}
   \end{subfigure}

   \caption{Empirical probability distributions $\hat{f}_{\boldsymbol{s}}$ [(a) and (d)], $\hat{f}^{27}_{\boldsymbol{a}}$ [(b) and (e)], and $\hat{f}_{\boldsymbol{g}}$ [(c) and (f)] for different $m$.}
   \label{fig:Emper-Proba-Mass-PD-PC}
\end{figure*}

In this section, we evaluate empirically the trade-offs among distortion $D$, rate $R$, and inference accuracy $A$ which are quantified by $\hat D$, $\hat R$, and $\hat A$, respectively. Note that the subscripts and the superscripts in the notation of those entities as we have presented in \S~\ref{subsec:Training-and-Testing} are dropped whenever needed for notational convenience. However, the suitable subscripts and/or superscripts are clear from context. The communication between \gls{tx} and \gls{rx} is assumed to be \emph{perfect} [i.e., error-free] in the simulation considered in \S~\ref{subsec:Trade-offs-between-Inference-Accuracy-and-Distortion} and \S~\ref{subsec:Trade-offs-between-Inference-Accuracy-and-Semantic-Rate}. Nevertheless, the simulation considered in \S~\ref{subsection:Exploiting-Reduction-in-Rates} assumes imperfect communication between \gls{tx} and \gls{rx}.

\subsubsection{Trade-offs between Distortion
and Rate}\label{subsec:Trade-offs-between-Semantic-Distortion-and-Rate}

Figure~\ref{fig:Trade-offs-between-Semantic-Distortion-and-Rate} shows the trade-off curves between distortion and rate. In the case of \gls{ae}-latent representations, the latent dimensions $2$ and $54$ are considered, i.e., $d=1$ and $d=27$. The points of the trade-off curves are obtained by changing the bins per dimension $m$ from $10$ to $27$. 

Figure~\ref{fig:Trade-offs-between-Semantic-Distortion-and-Rate} shows that there is a clear trade-off between distortion $D_{\textrm{MSE}}$ and the rate.
Results further show that \gls{pd} semantics always outperform \gls{ae}-latent representations and raw data. For example, in the case of \gls{pd} semantics, roughly a rate of $45$ [bits/object] is needed to achieve an \gls{mse} of $0.1$ while the \gls{ae}-latent representations and raw require a rate of several hundred [bits/object], roughly $20$ times more bits. Moreover, the seemingly constant gap between the curve associated with \gls{pd} semantics and other curves in Figure~\ref{fig:Trade-offs-between-Semantic-Distortion-and-Rate} indicates that the number of bits required by the others for achieving an \gls{mse} specification is simply a constant factor [e.g., $\approx20=13$dB] times the bits requirement of the case of \gls{pd} semantics. For example, to achieve an \gls{mse} distortion requirement of $0.02$, rate requirements are $250$ [bits/object] and $5000$ [bits/object] for \gls{pd} semantics and raw data, respectively. Results further show that we receive only a slight benefit in terms of \gls{mse} and rates if \gls{ae}-latent representations are used instead of raw data. Moreover,  the smaller the latent dimension, the better the performance of the representations. This is because the trade-off curve corresponding to the smaller $d$ (e.g., $d=1$) lies below the one corresponding to the larger $d$ (e.g., $d=27$) even though the differences are not significant.

To get some insight into why the rate requirements of \gls{pd} semantics are order of magnitude lower than that of both \gls{ae}-latent representations and raw data, let us plot the empirical probability distributions $\hat{f}_{\boldsymbol{s}}$, $\hat{f}^{d}_{\boldsymbol{a}}$, and $\hat{f}_{\boldsymbol{g}}$. Figure~\ref{fig:Emper-Proba-Mass-PD-PC} depicts empirically computed distributions $\hat{f}_{\boldsymbol{s}}$, $\hat{f}^{d}_{\boldsymbol{a}}$ and $\hat{f}_{\boldsymbol{g}}$ for $m \in \{10,24\}$ according to~\eqref{eq:empirical-probability-distribition}. The empirical distributions show that the probability masses associated with \gls{pd} semantics are concentrated on a smaller subset of possible outcomes while many with zero or almost zero probability. In contrast, the distributions corresponding to \gls{ae}-latent representations and raw data are not sparse and the probability masses of the outcomes are more or less comparable in general. Therefore, the results suggest that the uncertainty of \gls{pd} semantics is lower than that of \gls{ae}-latent representations and raw data. This is intuitively expected since \glspl{pd} encode structural properties of the underlying raw data, unlike \gls{ae} or the raw data itself. As a direct consequence, we see that the rate requirements for transmitting \gls{pd} semantics are significantly lower than the \gls{ae} and raw representations.

The preceding observations give important insights into engineering designs to make systems more robust against noise, interference, and fading in practical communication systems. More specifically, the additional bits required in the case of \gls{ae}-latent representations and raw data can be used in a system based on \gls{pd} semantics for improving the corresponding \gls{mse}. On the other hand, for a given \gls{mse} specification, the additional bits required in the case of \gls{ae}-latent representations and raw data can directly be used in systems that rely on \gls{pd} semantics to enable the possibility of implementing error detection/correction capabilities yielding more \emph{robustness}.
Insights into such advantages of \gls{pd} semantics over \gls{ae}-latent representations and raw data under an imperfect communication channel are given in~\S~\ref{subsection:Exploiting-Reduction-in-Rates}.



\subsubsection{Trade-offs between Inference Accuracy and Distortion}\label{subsec:Trade-offs-between-Inference-Accuracy-and-Distortion}

Figure~\ref{fig:Trade-offs-between-Inference-Accuracy-and-Distortion} shows the empirical trade-off curves between inference accuracy and \gls{mse} distortion. The thick curves depict the average results that correspond to trained \gls{nn} models for classification as discussed in \S~\ref{subsec:Training-and-Testing} [\cf~\eqref{eq:Av-Inf-Acuracy}] and the shaded regions highlight their variability. We also plot the trend in thin solid lines as a guideline obtained with function fitting.

Results show that the inference accuracies of models based on raw data tend to deteriorate as \gls{mse} distortion increases. In contrast, the inference accuracies of models based on \gls{pd} semantics and \gls{ae}-latent representations remain approximately constant, despite the level of \gls{mse}, over the considered range. This is indeed an interesting observation from a system design perspective because it demonstrates the advantages of semantics over raw data. As such, this empirical observation suggests that \gls{nn} models for classification trained with \gls{pd} semantics and \gls{ae}-latent representations are more robust than those based on raw data. Results further show that the smaller the latent dimension of \gls{ae}, the worse the inference accuracy. For example, \gls{ae}-latent representations with $d=1$ are clearly inferior to the \gls{pd} semantic case. However, a higher choice for $d$ can enable \gls{ae}-latent representations to perform better than \gls{pd} semantics.

\subsubsection{Trade-offs between Inference Accuracy and Rate}\label{subsec:Trade-offs-between-Inference-Accuracy-and-Semantic-Rate}

\begin{figure}[!t] 
 \centering
    \includegraphics[width=0.48\textwidth]{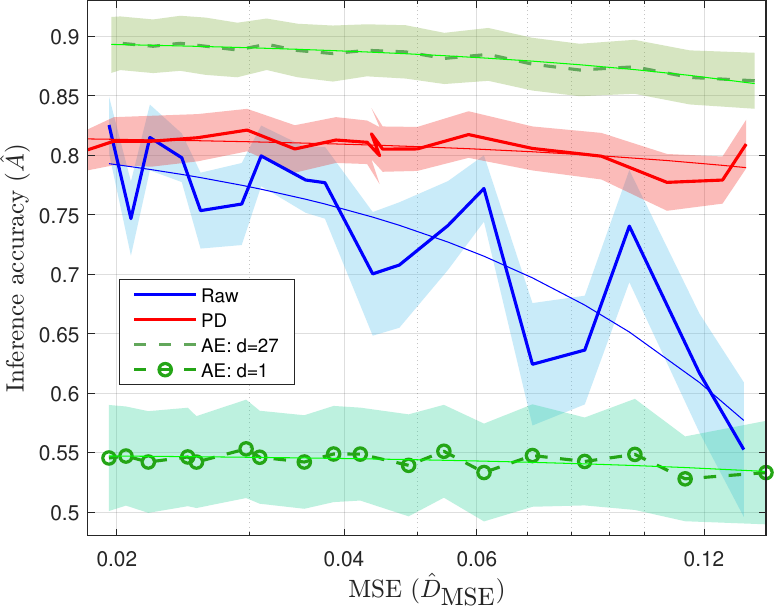}
   \caption{Empirical trade-off between $\hat A$ and  $\hat D_{\textrm{MSE}}$.}
   \label{fig:Trade-offs-between-Inference-Accuracy-and-Distortion}
\end{figure}

It is of practical interest to know how the inference accuracy is changed with rate specifications. Figure~\ref{fig:Trade-offs-between-Inference-Accuracy-and-Rate} shows the trade-off curves between the inference accuracy and the rate. The behavior of the curves is directly dictated by the preceding trade-offs experienced between the distortion and the rate, and those between the inference accuracy and the distortion, \cf~Figure~\ref{fig:Trade-offs-between-Semantic-Distortion-and-Rate} and \ref{fig:Trade-offs-between-Inference-Accuracy-and-Distortion}. 

At first glance, results show that stringent rate specification can only make the inference accuracy worse for \gls{nn} models based on raw data, \cf~blue solid curve. A similar behavior is also evident in the case of \gls{ae}-latent representations, especially with $d=27$, even though the deterioration of inference accuracy as the rate decreases is not as significant as in the case of raw data, \cf~thick green dashed curve. In contrast, the changes in the rates have an unnoticeable effect on the inference accuracy when the \gls{nn} models are trained on \gls{pd} semantics, \cf~red solid curves. It is worth highlighting that \gls{pd} semantics can yield inference accuracies over $80\%$ even with very low rate specifications such as $30.58$ [bits/object]. Reasons for such observations are evidenced by the characteristics of probability density $\hat{f}_{\boldsymbol{s}}$ as depicted in Figure~\ref{fig:Emper-Proba-Mass-PD-PC}.




\subsubsection{Exploiting Low \gls{pd} Rates under Imperfect Communications}\label{subsection:Exploiting-Reduction-in-Rates}

\begin{figure}[!t] 
 \centering
    \includegraphics[width=0.48\textwidth]{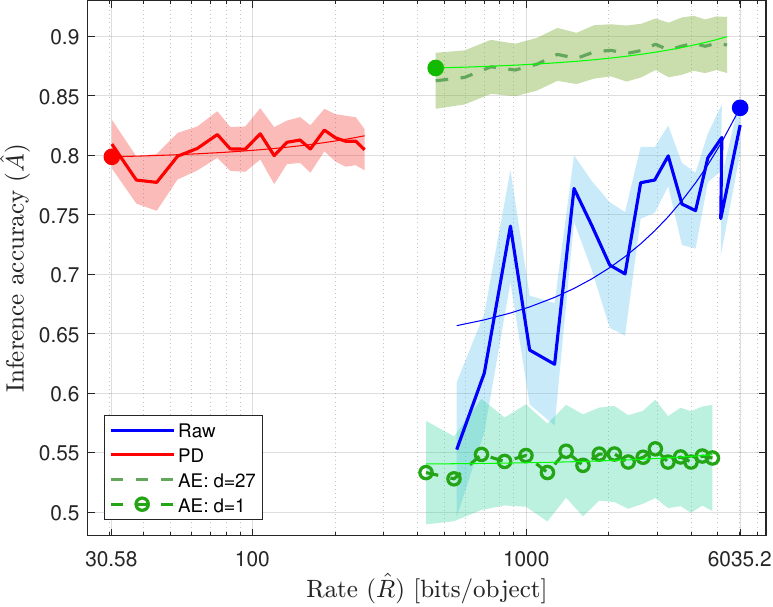}
   \caption{Empirical trade-off between $\hat A$ and $\hat R$.}
   \label{fig:Trade-offs-between-Inference-Accuracy-and-Rate}
\end{figure}

Note that the curves depicted in Figure~\ref{fig:Trade-offs-between-Inference-Accuracy-and-Distortion} and \ref{fig:Trade-offs-between-Inference-Accuracy-and-Rate} correspond to settings where the communication between \gls{tx} and \gls{rx} is perfect. In contrast, this section considers an imperfect communication between \gls{tx} and \gls{rx}. As a representative channel model, the \gls{bsc}~\cite[\S~7.1.4]{cover2012elements} is chosen, \cf~stage~$5$ of Figure~\ref{fig:Sys-Model}.  \gls{bsc} complements the input binary symbols with probability $\alpha\in[0,0.5)$~\footnote{The error-free communication channel is a particular case derived when $\alpha=0$ [\cf~\S~\ref{subsec:Trade-offs-between-Inference-Accuracy-and-Distortion} and \S~\ref{subsec:Trade-offs-between-Inference-Accuracy-and-Semantic-Rate}].}, which is referred to as the crossover probability. In this respect, the stage~$4$ of Figure~\ref{fig:Sys-Model} is implemented in two steps. First, the quantized \gls{pd} semantics are source-encoded using Huffman codes~\cite[\S~5.8]{cover2012elements} based on probabilities $p_k$, $k=1,\ldots,m^2$ [\cf~\eqref{eq:Probability-s-in-Rk}, ~\eqref{eq:empirical-probability-distribition}]. Then the resulting source-coded bit stream is channel-coded using $(\bar{n}, \bar{k}, \bar{t})$-\gls{bch} codes~\cite[\S~6]{Shu-Lin-etal}. Note that, given an $(\bar{n},\bar{k},\bar{t})$-\gls{bch} code, $\bar{n}$ denotes the codeword length and $\bar{k}$ denotes the length of binary information digits. The parameter~$\bar{t}$ denotes the error correction capability in the sense that the code is capable of correcting all the error patterns of $\bar{t}$ or exhibiting fewer errors.

\begin{figure}[!t] 
 \centering
    \includegraphics[width=0.48\textwidth]{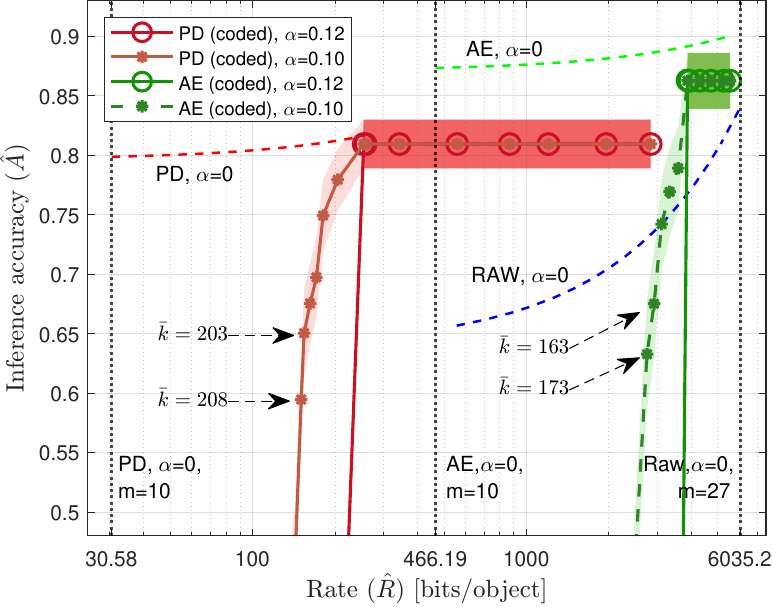}
   \caption{Empirical trade-off between $\hat A$ and  $\hat R$ (coded). For comparison, the trends of $\hat A$ versus rate $\hat R$ under perfect communication setting are reported as thin dashed curves.}
   \label{fig:Trade-offs-between-Inference-Imperfec-Comm}
\end{figure}

Before we give any empirical results under imperfect communication channel conditions to compare and contrast \gls{pd} semantic-based systems with \gls{ae}-latent representations or raw data-based systems, let us point out meaningful resource settings to provide a fair comparison.
First, for \gls{ae}, we consider the case of $d=27$ which yields a better inference accuracy. As we noticed in \S~\ref{subsec:Trade-offs-between-Inference-Accuracy-and-Semantic-Rate}, the inference accuracy of \gls{nn} models based on \gls{pd} semantics and \gls{ae}-latent representations remain almost unchanged over the considered range of rates when the communication is perfect, \cf~solid thin red and solid thin green curves in Figure~\ref{fig:Trade-offs-between-Inference-Accuracy-and-Rate}. As a result of this robustness of \gls{pd} semantics and \gls{ae}-latent representations, designing systems with a coarser quantization, such as $m=10$, is more suggestive since they account for lower rates. For example, note that the solid red circle and the solid green circle depicted in Figure~\ref{fig:Trade-offs-between-Inference-Accuracy-and-Rate}  correspond to \emph{self-information} values of 
\begin{equation*}
    r_{\textrm{PD}}=30.58 \quad  \quad \textrm{and} \quad  r_{\textrm{AE}}=466.19 \quad \textrm{[bits/object]},
\end{equation*}
respectively on average. On the other hand, if raw data are used in a perfect communication setting, unlike \gls{pd} semantics, they cannot achieve a similar inference accuracy with a coarser quantization level. Consequently, they have to rely on finer quantization values that account for higher self-information [e.g., $m=27$], \cf~solid blue circle, Figure~ \ref{fig:Trade-offs-between-Inference-Accuracy-and-Rate}. More specifically, to achieve a better inference accuracy level comparable to \gls{pd} semantics with $r_{\textrm{PD}}$ [bits/object], the raw data needs to operate with $m=27$ which corresponds to a rate of 
\begin{equation*}
    r_{\textrm{Raw}}=6035.20 \quad \textrm{[bits/object]}
\end{equation*}
on average. These rate limits are depicted by using dotted vertical lines in Figure~\ref{fig:Trade-offs-between-Inference-Imperfec-Comm}. Thus, from an engineering point of view, the preceding observation can be used as the key to boosting the performance of \gls{pd} semantic-based systems under imperfect communication settings. As such, in the~sequel, we consider a resource setting where no more than $r_{\textrm{Raw}}$ [bits/object] is used~\footnote{This is the resource requirement of the raw data-based systems to achieve the highest possible inference accuracy in our simulations.}.

A simple calculation suggests that as long as quantized \gls{pd} semantics is source encoded with an optimal Huffman code~\footnote{The source rates [\cf~\eqref{eq:Probability-s-in-Rk}, \eqref{eq:Entropy-of-quantizer-q}] are based on the entropy and thus, we rely on an optimal source code such as Huffman to perform source encoding.} followed by an $(\bar{n},\bar{k},\bar{t})$-\gls{bch} code chosen such that
\begin{equation}\label{eq:n-k-choice-for-BCH}
  \bar{k} \ r_{\textrm{Raw}}  \geq \bar{n} \ 
 r_{\textrm{PD}},
\end{equation}
the average length of the bit sequence $\boldsymbol{u}$ per object always remains  within the limit $r_{\textrm{Raw}}$~[bits/object]. In our experiments, we let $\bar{n}=1023$. Therefore, from \eqref{eq:n-k-choice-for-BCH}, we have $\bar{k}\geq 5.18$. Thus, all $(1023, \bar{k}, \bar{t})$-\gls{bch} codes with $\bar{k}\geq 6$ operate within the resource limit of $r_{\textrm{Raw}}$~[bits/object]. For example, suppose $\bar{k}=208$, i.e., the corresponding \gls{bch} code is $(1023, 208, 115)$ \cite[Table~6.1]{Shu-Lin-etal}. In this case, the actual communication rate requirement to transmit $r_{\textrm{PD}}=30.58$ information bits is simply $150.41$ bits which is indeed less than $r_{\textrm{Raw}}$. On the other hand, the coded \gls{pd} semantics become robust under imperfect communication settings since $(1023,208,115)$-\gls{bch} code is capable of correcting all error patterns of $115$ or fewer~errors. As such, the key is to encode quantized \gls{pd} semantics by using $(1023, \bar{k}, \bar{t})$-\gls{bch} codes whose $\bar{k}$ conforms to \eqref{eq:n-k-choice-for-BCH} with the hope of being more robust against channel impairments.
By replacing $r_{\textrm{PD}}$ in \eqref{eq:n-k-choice-for-BCH} by $r_{\textrm{AE}}$, $(1023, \bar{k}, \bar{t})$-\gls{bch} codes compatible with \gls{ae}-latent representations that  operate within the resource limit $r_{\textrm{Raw}}$~[bits/object] can be computed similarly.

Figure~\ref{fig:Trade-offs-between-Inference-Imperfec-Comm} shows the inference accuracy versus coded information rate requirements for quantized \gls{pd} semantics and quantized \gls{ae}-latent representations with $(1023, \bar{k}, \bar{t})$-\gls{bch} codes, where the bins per dimension $m=10$. Compatible \gls{bch} codes used with \gls{pd} semnatics correspond to $\bar{k}=208, 203, 193, 183, 173, 153, 123, 121, 91, 56, 36, 26, 16$, and  $11$. On the other hand, compatible \gls{bch} codes used with \gls{ae}-latent representations  correspond to $\bar{k}=173, 163, 153, 143, 133, 123, 111, 101, 91$, and $86$.
Simulations are conducted for two \gls{bsc} conditions. In particular, case~$1$ considers a \gls{bsc} with $\alpha=0.12$ and case~$2$ considers a \gls{bsc} with $\alpha=0.1$. 

Results show that the best inference accuracy levels of perfect communication settings are also attainable even under the imperfect communication settings when \gls{pd} semantics are coded with $(1023, \bar{k}, \bar{t})$-\gls{bch} codes.
We emphasize that this is achieved without utilizing no more than $r_{\textrm{Raw}}$ [bits/object]. For example, with coded \gls{pd} semantics, for both $\alpha=0.1$ and $\alpha=0.12$, the best inference accuracy is achieved by using $(1023, \bar{k}, \bar{t})$-\gls{bch} codes with $\bar{k}= 123, 121, 91, 56, 36, 26, 16$, and $11$ that correspond to coded information rates of $254.35, 258.55, 343.79, 558.66, 869.02, 1203.26, 1955.30$, and $2844.08$ [bits/object], respectively. Of course, the code $(1023, 123, 170)$ with the lowest resource requirement (i.e., highest code rate), is preferable among those. It is worth pointing out that, such a choice corresponds to more than $23$-fold communication resource reduction compared to $r_{\textrm{Raw}}$ [bits/object], even under imperfect communication. Thus, results suggest that when a system based on raw data under perfect communication settings infers one object, a system based on \gls{pd} semantics will infer $23$ objects while maintaining the same order of inference accuracy even under imperfect communication. 

Coded \gls{ae}-latent representations can also yield channel error-free inference accuracy levels with some $(1023, \bar{k}, \bar{t})$-\gls{bch} codes. For example, \gls{bch} codes with $\bar{k}=133, 123, 111, 101, 91$, and $86$ that correspond to coded information rates of $3877.38, 3941.47, 4296.55, 4721.96, 5240.85$, and $5545.55$ [bits/object], yield the best inference accuracy under both \gls{bch} channel conditions $\alpha=0.1$ and $\alpha=0.12$. Therefore, results show that the rate requirements in coded \gls{ae}-latent representations are significantly higher compared to the rate requirements of coded \gls{pd} semantics. For example, the minimum rate requirement of coded \gls{pd} semantics to operate at an inference accuracy of more than $80$\% is, roughly, $15$ times as small as the rate requirement of coded \gls{ae}-latent representations. As such, in practice, the choice of \gls{pd} semantics enables not only withstanding the imperfect communication impairments effectively but also the improvements in the \emph{latency} compared to the choices of both \gls{ae}-latent representations and raw data.































\begin{figure*}[!t] 
 \centering
   \begin{subfigure}{0.326\textwidth}
     \includegraphics[width=\linewidth]{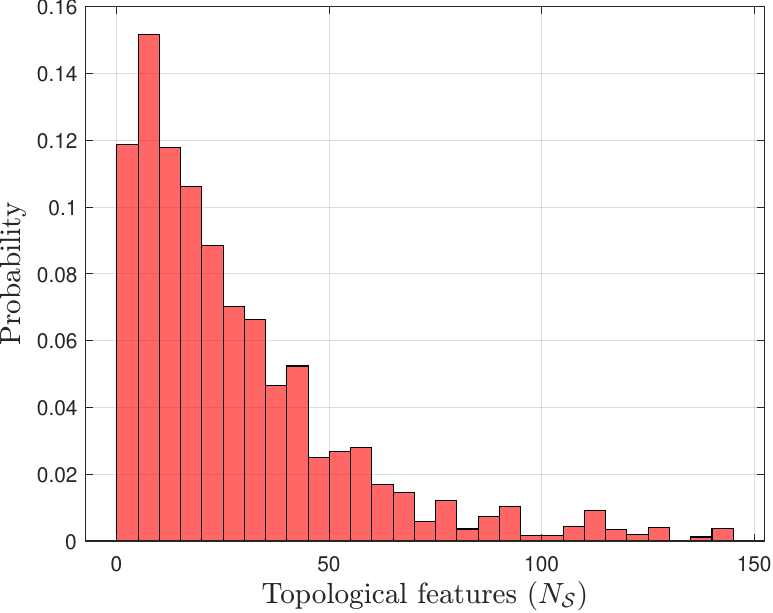}
       \caption{}
       \label{fig:No-Featu-PD}
   \end{subfigure}
\hfill 
   \begin{subfigure}{0.326\textwidth}
       \includegraphics[width=\linewidth,height=5cm]{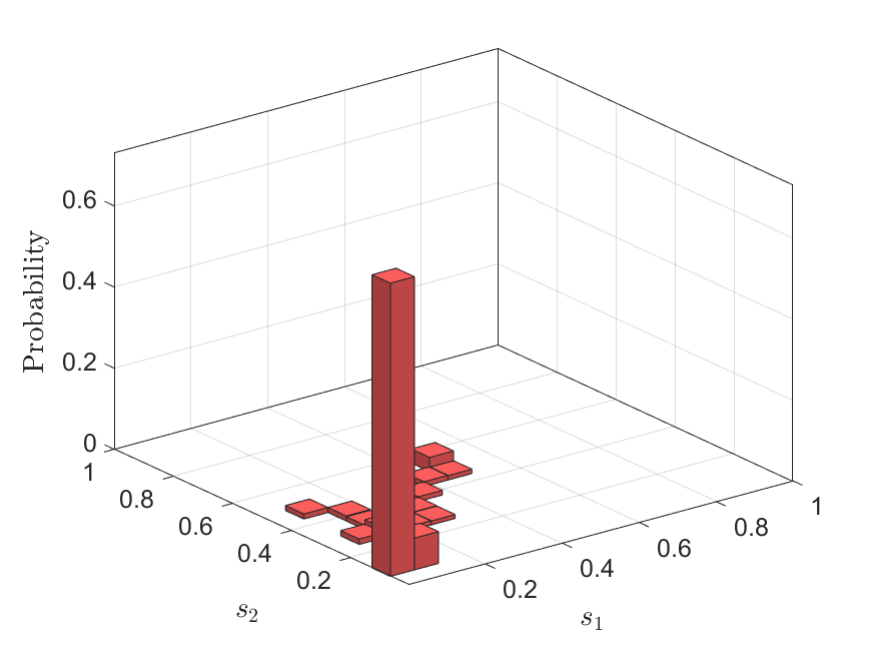}
       \caption{}
       \label{fig:Cond-Dist-PD1}
   \end{subfigure}
\hfill 
   \begin{subfigure}{0.326\textwidth}
       \includegraphics[width=\linewidth,height=5cm]{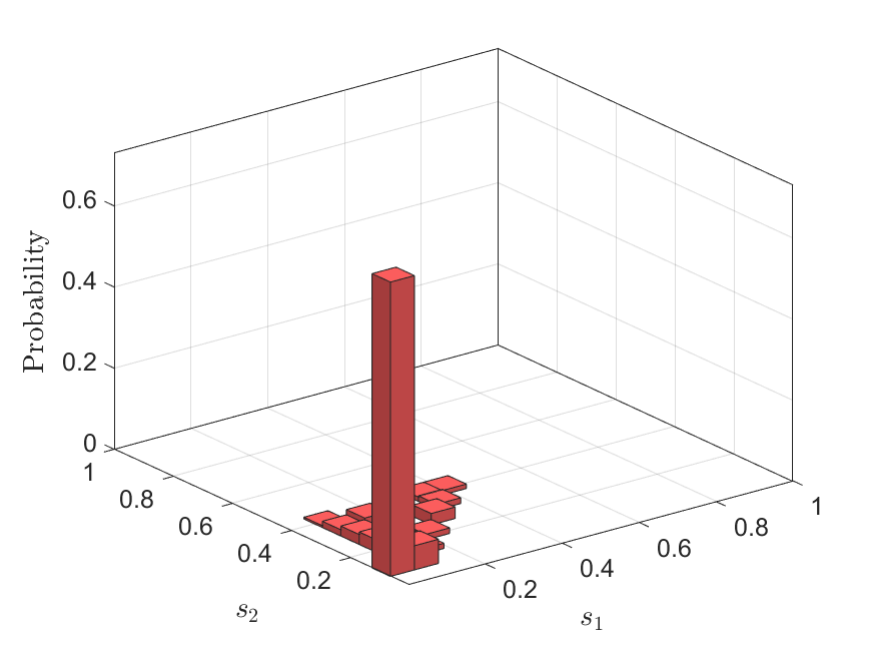}
       \caption{}
       \label{fig:Cond-Dist-PD2}
   \end{subfigure}

 \begin{subfigure}{0.326\textwidth}
     \includegraphics[width=\linewidth]{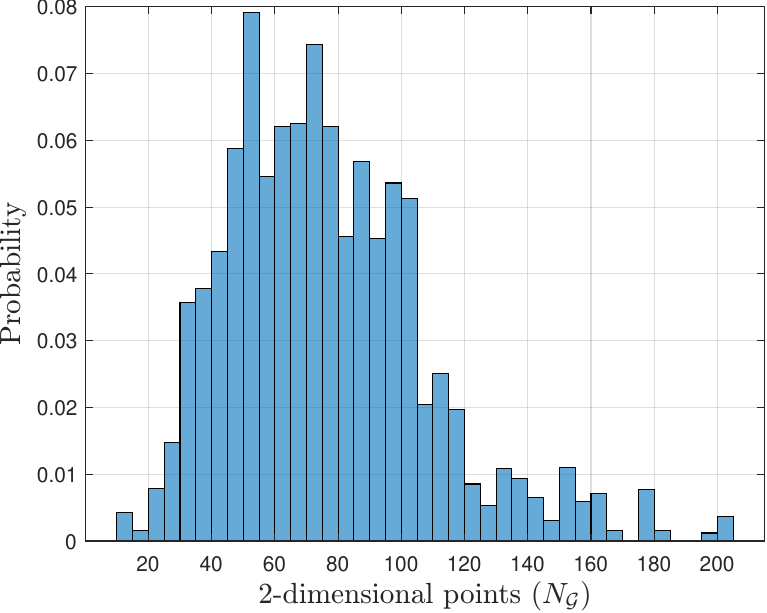}
       \caption{}
       \label{fig:No-Featu-PC}
   \end{subfigure}
\hfill 
   \begin{subfigure}{0.326\textwidth}
       \includegraphics[width=\linewidth,height=5cm]{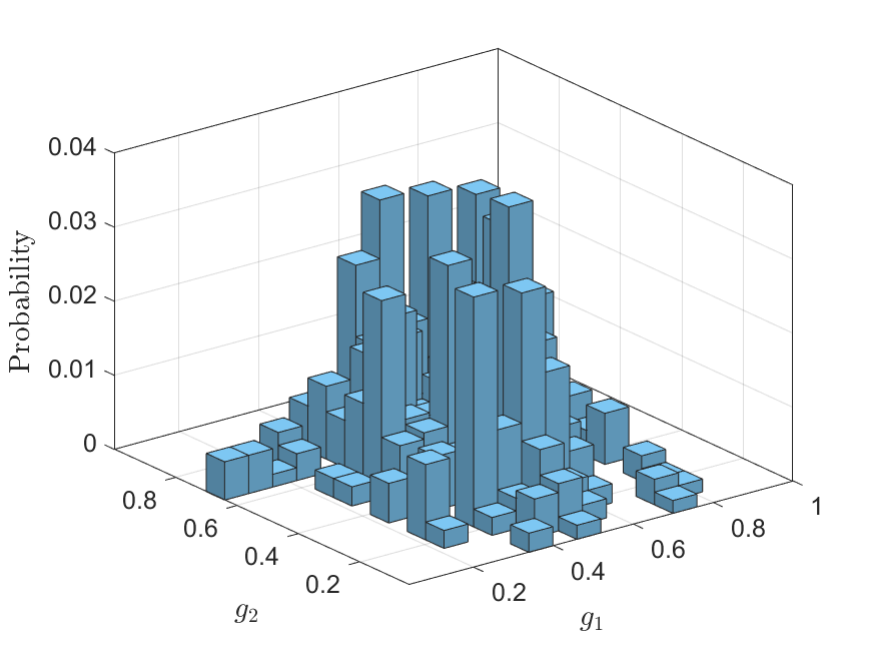}
       \caption{}
       \label{fig:Cond-Dist-PC1}
   \end{subfigure}
\hfill 
   \begin{subfigure}{0.326\textwidth}
       \includegraphics[width=\linewidth,height=5cm]{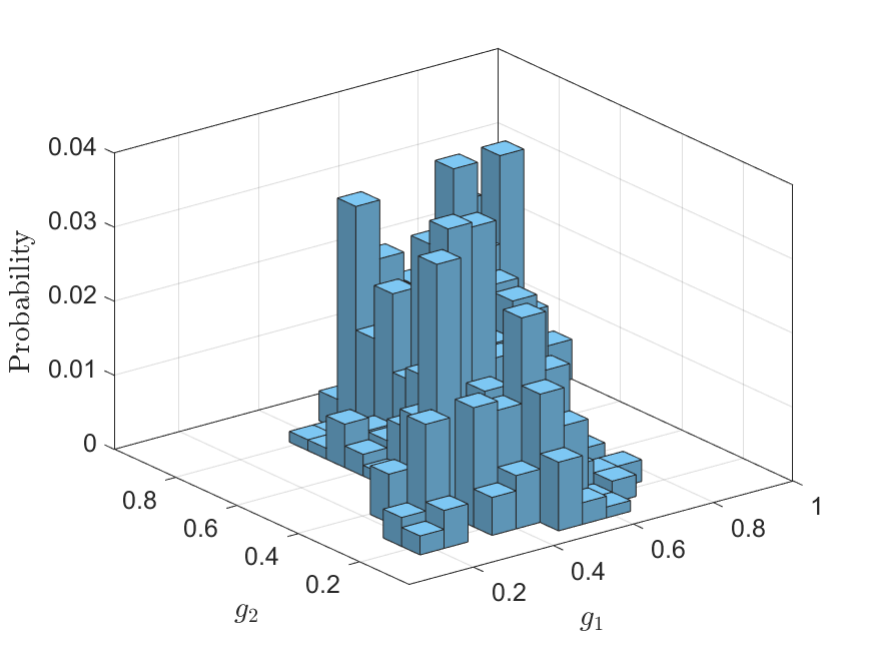}
       \caption{}
       \label{fig:Cond-Dist-PC2}
   \end{subfigure}
   
   \caption{Empirical probability distributions where $\boldsymbol{s}_1,\ldots,\boldsymbol{s}_{N_{\mathcal{S}}}$ (and $\boldsymbol{g}_1,\ldots,\boldsymbol{g}_{N_{\mathcal{G}}}$) are conditionally \gls{iid} given $N_{\mathcal{S}}$ (and $N_{\mathcal{G}}$) according to $f_{\boldsymbol{s}|N_{\mathcal{S}}}$ \big(and $f_{\boldsymbol{g}|N_{\mathcal{G}}}$\big): (a) $\hat{f}_{N_{\mathcal{S}}}$, (b) $\hat{f}_{\boldsymbol{s}|N_{\mathcal{S}}=6}$ for $m=16$, (c) $\hat{f}_{\boldsymbol{s}|N_{\mathcal{S}}=10}$ for $m=16$, (d) $\hat{f}_{N_{\mathcal{G}}}$, (e) $\hat{f}_{\boldsymbol{g}|N_{\mathcal{G}}=50}$ for $m=16$, (f) $\hat{f}_{\boldsymbol{g}|N_{\mathcal{G}}=100}$ for $m=16$.}
   \label{fig:Emperical-General-Probability-Distributions}
\end{figure*}

\glsreset{pd}
\glsreset{ml}

\section{Conclusion}\label{sec:Conclusion}

In this work, we explored the advantages of \glspl{pd} over raw point cloud data as structural semantics for a point-to-point communication setting. The transmitter (e.g., sensor) used \gls{pd} semantics in its transmissions and the receiver (e.g., central node) carried out inference using the received semantics. We developed qualitative definitions for distortion and rate of \gls{pd} semantics and quantitatively characterized the trade-offs among the distortion, the rate, and the inference accuracy at the receiver node. Our empirical results demonstrated that \gls{pd} semantics leads to more effective use of transmission channels, enhanced degrees of freedom for incorporating error detection and correction capabilities, and improved robustness to communication channel imperfections. Results further highlighted the potential of using structural semantics in goal-oriented communication settings ensuring high efficiency and robustness of communication systems, where intrinsic channel impairments, such as noise, interference, and fading are inevitable.





\appendices


\section{A General Distortion Definition for  $\mathcal{S}$}~\label{App:Dist-over-S}
Our second notion of semantic distortion is inspired by the bottleneck distance~\eqref{eq:Bottleneck}. However, the map $\eta$ in~\eqref{eq:Bottleneck} is restricted to be the quantization map $Q^m$ given in \eqref{eq:Quantization-Map}. More specifically, the distortion of $\mathcal{S}$, denoted $D_{\textrm{BN}}^{\textrm{sem}}$, is given by
\begin{equation}\label{eq:eq:Semantic-Distortion-BN}\nonumber
    D_{\textrm{BN}}^{\textrm{sem}} = \mathbb{E}_{f_{N_{\mathcal{S}}}} \mathbb{E}_{f_{\boldsymbol{s}_1,\ldots,  \boldsymbol{s}_{N_{\mathcal{S}}}|N_{\mathcal{S}}}} \left[\max_{i\in \{1,2,\ldots,N_{\mathcal{S}}\}} ||\boldsymbol{s}_i-q^m(\boldsymbol{s_i})||_{\infty}\right], 
\end{equation}
where the expectation $\mathbb{E}_{f_{\boldsymbol{s}_1,\ldots,\boldsymbol{s}_{N_{\mathcal{S}}}|N_{\mathcal{S}}}}$ is taken with respect to the  conditional  distribution $f_{\boldsymbol{s}_1,\ldots,\boldsymbol{s}_{N_{\mathcal{S}}}|N_{\mathcal{S}}}$ and expectation $\mathbb{E}_{f_{N_{\mathcal{S}}}}$ is taken with respect to the distribution  $f_{N_{\mathcal{S}}}$. A similar notion of distortion can also be constructed for raw data based on $f_{\boldsymbol{g}_1,\ldots,\boldsymbol{g}_{N_{\mathcal{S}}}|N_{\mathcal{G}}}$ and $f_{N_{\mathcal{G}}}$.

The distributions $f_{\boldsymbol{s}_1,\ldots,\boldsymbol{s}_{N_{\mathcal{S}}}|N_{\mathcal{S}}}$, $f_{N_{\mathcal{S}}}$, $f_{\boldsymbol{g}_1,\ldots,\boldsymbol{g}_{N_{\mathcal{G}}}|N_{\mathcal{S}}}$, and $f_{N_{\mathcal{G}}}$ are not available in practice and therefore, one must rely on numerical approaches to yield estimates. Assuming $\boldsymbol{s}_1,\ldots,\boldsymbol{s}_{N_{\mathcal{S}}}$ (and $\boldsymbol{g}_1,\ldots,\boldsymbol{g}_{N_{\mathcal{G}}}$) are conditionally \gls{iid} given $N_{\mathcal{S}}$ (and $N_{\mathcal{G}}$) according to some conditional distribution $f_{\boldsymbol{s}|N_{\mathcal{S}}}$ \big(and $f_{\boldsymbol{g}|N_{\mathcal{G}}}$\big), Figure~\ref{fig:Emperical-General-Probability-Distributions} illustrates the related empirical probability distributions $\hat{f}_{N_{\mathcal{S}}}, \hat{f}_{\boldsymbol{s}|N_{\mathcal{S}}}, \hat{f}_{N_{\mathcal{G}}}$, and $\hat{f}_{\boldsymbol{g}|N_{\mathcal{G}}}$ for some choices of $m$ based on what the corresponding distortions can numerically be evaluated.

Note that the sparsity characteristics of the empirical conditional distributions are very similar to those in Figure~\ref{fig:Emper-Proba-Mass-PD-PC}. That is the probability masses of the distributions associated with \gls{pd} semantics are concentrated on a few outcomes and zero or almost zero for most of the outcomes, unlike the the probability masses of conditional distributions that correspond to raw data. Thus, the results suggest that the numerical evaluations of trade-off curves with the general distortion definition behave similarly to those with $D_{\textrm{MSE}}^{\textrm{sem}}$.

\section{Parameters for the Empirical Simulations}~\label{App:Para-NN}
 
For computing \gls{pd} $\mathcal{S}_i$ from a given raw data $\mathcal{G}_i$, we  employed the \gls{vr} filtration with a positive scalar $\gamma_t=16$, \cf~\eqref{eq:VR-Filtration}. Vectorized representations of the \glspl{pd} were then generated using \texttt{PersLay}, with its parameters defined as follows:
$\texttt{op}:=\texttt{top2}$, $w:=\texttt{PowerPerslayWeight}$, and $\phi:=\texttt{TentPerslayPhi}$, \cf~\eqref{eq:PersLay}.  These computations were carried out using the GUDHI library~\cite{gudhi:urm}.


Our empirical simulations employed \gls{nn} models, with their parameters summarized in Table~\ref{table:sim_param}. Moreover, the \glspl{ae} were trained on raw data $\{\mathcal{G}_i\}_{i\in\mathcal{N}^{\textrm{train}}_{26}}$ using the binary cross-entropy loss function. The classifiers were trained with the categorical cross-entropy loss function, and all models were optimized using the ADAM optimizer with a learning rate of $0.001$.

\begin{table}[!htb]
\centering
\def\arraystretch{1.25}%
\caption{A summary of the \gls{nn} models.}
    \centering
    \begin{tabular}{|c|c|c|c|}
     \hline
        Purpose & Model  &  $\#$ of parameters & Library\\
       \hline
 \multirow{2}{*}{Training encoders}      & $\texttt{Enc}^{27}$ & $442566$ & PyTorch~\cite{paszke2019pytorch}\\
        &  $\texttt{Enc}^{1}$ & $435858$ & PyTorch\\
        \hline
 \multirow{4}{*}{Training classifiers}       & $\texttt{PD}_t$ & $201747$ & TensorFlow~\cite{abadi2015tensorflow}\\
        & $\texttt{Raw}_t$ & $201731$ & PyTorch\\
        & $\texttt{AE}_t^{27}$ & $2339$ & PyTorch\\
        & $\texttt{AE}_t^{1}$ & $27$ & PyTorch\\
        \hline
    \end{tabular}
    \label{table:sim_param}
\end{table}

\bibliographystyle{IEEEtran}
\bibliography{References.bib}


    

\end{document}